\documentclass[sigconf]{acmart}

\usepackage{etoolbox}
\makeatletter
\patchcmd{\maketitle}{\includegraphics[height=5ex]{doclicense-CC-\ACM@cc@type-88x31}}{}{}{}
\makeatother

\makeatletter
\def\ACM@cc@type{by}
\def\ACM@cc@version{4.0}

\def\@doclicense@\@empty
\let\doclicenseThis\relax
\makeatother

\AtBeginDocument{%
  \providecommand\BibTeX{{%
    Bib\TeX}}}

\copyrightyear{2026}
\acmYear{2026}
\setcopyright{cc}
\setcctype{by}
\acmConference[KDD '26]{Proceedings of the 32nd ACM SIGKDD Conference on Knowledge Discovery and Data Mining V.2}{August 09--13, 2026}{Jeju Island, Republic of Korea}
\acmBooktitle{Proceedings of the 32nd ACM SIGKDD Conference on Knowledge Discovery and Data Mining V.2 (KDD '26), August 09--13, 2026, Jeju Island, Republic of Korea}
\acmDOI{10.1145/3770855.3818506}
\acmISBN{979-8-4007-2259-2/2026/08}

\usepackage[linesnumbered,ruled,vlined]{algorithm2e}
\usepackage{xcolor} 
\usepackage{enumitem}
\usepackage{multirow}
\usepackage[frozencache, cachedir=.]{minted}
\SetCommentSty{mycommfont}

\begin{document}

\title{Vectorizing the Trie: Efficient Constrained Decoding for LLM-based Generative Retrieval on Accelerators}


\settopmatter{authorsperrow=4} 


\author{Zhengyang Su}
\authornote{These authors contributed equally to this work.}
\affiliation{%
  \institution{Google}
  \city{Mountain View, CA}
  \country{USA}
}
\email{susteven@google.com}

\author{Isay Katsman}
\authornotemark[1]
\authornote{This work was performed while the author was participating in the Google Student Researcher Program at Google.}
\affiliation{%
  \institution{Yale University}
  \city{New Haven, CT}
  \country{USA}
}
\email{isay.katsman@yale.edu}

\author{Yueqi Wang}
\authornotemark[1]
\affiliation{%
  \institution{Google}
  \city{New York, NY}
  \country{USA}
}
\email{yueqiw@google.com}

\author{Ruining He}
\affiliation{%
  \institution{Google}
  \city{Mountain View, CA}
  \country{USA}
}
\email{ruininghe@google.com}

\author{Lukasz Heldt}
\affiliation{%
  \institution{Google}
  \city{Mountain View, CA}
  \country{USA}
}
\email{heldt@google.com}

\author{Raghunandan Keshavan}
\affiliation{%
  \institution{Google}
  \city{Mountain View, CA}
  \country{USA}
}
\email{hkraghunandan@google.com}

\author{Shao-Chuan Wang}
\affiliation{%
  \institution{Google}
  \city{Mountain View, CA}
  \country{USA}
}
\email{scwang@google.com}

\author{Xinyang Yi}
\affiliation{%
  \institution{Google}
  \city{Mountain View, CA}
  \country{USA}
}
\email{xinyang@google.com}

\author{Mingyan Gao}
\affiliation{%
  \institution{Google}
  \city{Mountain View, CA}
  \country{USA}
}
\email{mingyan@google.com}

\author{Onkar Dalal}
\affiliation{%
  \institution{Google}
  \city{Mountain View, CA}
  \country{USA}
}
\email{onkardalal@google.com}

\author{Lichan Hong}
\affiliation{%
  \institution{Google}
  \city{Mountain View, CA}
  \country{USA}
}
\email{lichan@google.com}

\author{Ed H. Chi}
\affiliation{%
  \institution{Google}
  \city{Mountain View, CA}
  \country{USA}
}
\email{edchi@google.com}

\author{Ningren Han}
\affiliation{%
  \institution{Google}
  \city{Mountain View, CA}
  \country{USA}
}
\email{peterhan@google.com}
\settopmatter{authorsperrow=4}



\renewcommand{\shortauthors}{Zhengyang Su et al.}

\begin{abstract}
    Generative retrieval has emerged as a powerful paradigm for LLM-based recommendation. However, industrial recommender systems often benefit from restricting the output space to a constrained subset of items based on business logic (e.g. enforcing content freshness or product category), which standard autoregressive decoding cannot natively support. Moreover, existing constrained decoding methods that make use of prefix trees (Tries) incur severe latency penalties on hardware accelerators (TPUs/GPUs). In this work, we introduce \textbf{STATIC} (\textbf{S}parse \textbf{T}ransition Matrix-\textbf{A}ccelerated \textbf{T}rie \textbf{I}ndex for \textbf{C}onstrained Decoding), an efficient and scalable constrained decoding technique designed specifically for high-throughput LLM-based generative retrieval on TPUs/GPUs. By flattening the prefix tree into a static Compressed Sparse Row (CSR) matrix, we transform irregular tree traversals into fully vectorized sparse matrix operations, unlocking massive efficiency gains on hardware accelerators.

We deploy STATIC on a large-scale industrial video recommendation platform serving billions of users. STATIC produces significant product metric impact with minimal latency overhead ($0.033$ ms per step and $0.25\%$ of inference time), achieving a $948\text{\small $\times$}$ speedup over a CPU trie implementation and a $47$--$1033 \text{\small $\times$}$ speedup over a hardware-accelerated binary-search baseline. Furthermore, the runtime overhead of STATIC remains extremely low across a wide range of practical configurations. To the best of our knowledge, STATIC enables the first production-scale deployment of strictly constrained generative retrieval. In addition, evaluation on academic benchmarks demonstrates that STATIC can considerably improve cold-start performance for generative retrieval. Our code is available at {\color{blue}\url{https://github.com/youtube/static-constraint-decoding}}.

\end{abstract}




\maketitle

\section{Introduction}
\label{sec:intro}

Recommender systems have long served as the core discovery engines powering massive online platforms like YouTube and Amazon \cite{Linden2003AmazoncomRI, Covington2016DeepNN}. While neural network techniques continue to improve these engines, this domain is undergoing a paradigm shift whereby embedding-based neural retrieval models that recommend candidates through approximate nearest neighbor search such as ScaNN \cite{Guo2019AcceleratingLI, Sun2024SOARII} are being superseded with LLM-based generative retrieval \cite{Rajput2023RecommenderSW, He2025PLUMAP, Deng2025OneRecUR, Yang2024UnifyingGA, Si2023GenerativeRW, zhou2026openonerectechnicalreport}. By representing each item as a sequence of discrete tokens (called Semantic IDs or SIDs) and training Transformers \cite{Vaswani2017AttentionIA} to autoregressively decode the Semantic ID tokens of the target items \cite{Rajput2023RecommenderSW}, generative retrieval bypasses key limitations of traditional dual (user, item)-encoders \cite{weller2026on} to capture deeper semantic relationships. Additionally, it obviates the need for an external nearest-neighbor indexing infrastructure \cite{Rajput2023RecommenderSW} and allows one to adapt pre-trained LLMs for recommendation tasks \cite{He2025PLUMAP, Liu2025OneRecThinkIR, zhou2026openonerectechnicalreport}. 

However, generative retrieval models as they were originally designed lack a critical feature: control over the generated output space, as required to enforce business logic across a variety of industry applications. The absence of this feature has not stopped these next generation recommendation models from becoming widespread in large-scale industry settings \cite{He2025PLUMAP, Deng2025OneRecUR, Zhou2025OneRecTR, Zhou2025OneRecV2TR}, but more fine-grained output space control 
would enable practitioners to restrict the output space based on business logic, such as enforcing freshness constraints (e.g. ``uploaded within the last day''), locality (e.g. regional recommendations), product categories (e.g. ``recommend summer clothes''), or inventory availability (e.g. ``in stock'' only). This makes it possible to train a single large model and deploy it across multiple use cases.

In a traditional recommendation setting, business logic is enforced by filtering items post-retrieval or constraining the search space a priori (e.g. by running a restricted ScaNN search \cite{Guo2019AcceleratingLI}). If a system requires items to be ``in stock'' or ``uploaded within 7 days,'' a boolean filter simply drops invalid candidates from the retrieved set or the ScaNN index. In generative retrieval, however, the ``retriever'' is a probabilistic autoregressive model that synthesizes Semantic IDs \cite{Rajput2023RecommenderSW}. Without intervention, an LLM will confidently generate SIDs for items that are out-of-stock, stale, or legally restricted. Relying on post-generation filtering is computationally wasteful; the model may spend its entire inference budget generating invalid items, resulting in zero valid recommendations. 

To address this issue, constraints must be enforced \textit{during} the LLM decoding process. The standard algorithmic solution is to implement constrained decoding with a prefix tree (trie), which masks invalid tokens at each step \cite{Ye2025EfficientAA}. While conceptually sound, we observe that naive pointer-chasing trie-based constraints are fundamentally hostile to hardware accelerators for two reasons:

\begin{enumerate}
    \item \textbf{Memory Latency} Pointer-based structures result in non-contiguous, random memory access patterns. This prevents memory coalescing, failing to utilize the High-Bandwidth Memory (HBM) burst capabilities of modern TPUs/GPUs \cite{Jouppi2017IndatacenterPA}. Furthermore, these irregular patterns nullify hardware prefetchers designed for linear streaming, leading to severe memory-bound latency and stalled execution units.
    \item \textbf{Compilation Incompatibility} Modern accelerator architectures (e.g. Google's XLA-reliant \cite{sabne2020xla} TPUs) require static computation graphs for ML compilation \cite{Li2020TheDL}. Naive trie implementations rely on data-dependent control flow and irregular linked structures, which are fundamentally incompatible with this paradigm. While accelerators support dynamic indexing into fixed-size buffers (e.g. vectorized gather), the irregular memory access and recursive branching of pointer-chasing preclude end-to-end ML compilation.
\end{enumerate}

In our preliminary experiments, a CPU-offloaded trie implementation increased inference time by $2\text{\small $\times$}$, rendering it unusable for our target per-decoding-step latency of $\leq\!10$ ms. More sophisticated existing approaches utilizing binary search also exhibited significant latency penalties \cite{Ye2025EfficientAA}. While alternative methods like Finite State Transducers (FSTs) are standard in natural language processing, they suffer from state explosion when applied to production-scale SID vocabularies (e.g. on the order of millions of items) and lack native TPU support \cite{Koo2024AutomatabasedCF}.

In this work, we introduce \textbf{STATIC} (\textbf{S}parse \textbf{T}ransition Matrix-\textbf{A}ccelerated \textbf{T}rie \textbf{I}ndex for \textbf{C}onstrained Decoding), a framework that recasts constrained decoding from a graph traversal problem into a series of vectorized sparse matrix operations. STATIC is designed to resolve the aforementioned issues and enable extremely fast and scalable constrained decoding for generative retrieval. Notably, our I/O complexity\footnote{Unlike traditional time or space complexity, I/O complexity \cite{Hong1981IOCT} on a TPU evaluates algorithmic efficiency by counting the number of costly data block transfers between off-chip High Bandwidth Memory (HBM) and on-chip SRAM.} is $O(1)$ with respect to constraint set size, in contrast with the logarithmic scaling from existing binary search-based methods.

Our contributions are as follows:
\begin{enumerate}[label=(\roman*)]
    \item We propose a method to flatten prefix tree-specified constraints into static Compressed Sparse Row (CSR) matrices, enabling $O(1)$ memory access overhead via coalesced reads for fast decoding constraint extraction.
    \item We design a branch-free decoding algorithm using dynamic slicing and mask arithmetic. This makes constrained decoding fully accelerator-native and eliminates host-device round-trips, thereby enabling massive efficiency gain.
    \item We deploy this system on an ultra-large-scale video recommendation platform (YouTube) for multiple product use cases. We present one specific setting in which we constrain an generative retrieval LLM to a pre-specified vocabulary of $20$ million fresh items, improving key online metrics with minimal latency overhead compared to baseline methods.
    \item We evaluate the scalability of STATIC and find that its latency remains extremely low across a wide range of constraint set and Semantic ID vocabulary sizes.
    \item We demonstrate how one can use constrained decoding to improve cold-start recommendation performance in a principled setting using Amazon Reviews \cite{He2016UpsAD} datasets.
\end{enumerate}

\section{Related Work}
\label{sec:related_work}

In the recommender systems field, the shift from embedding-based retrieval \cite{Guo2019AcceleratingLI} to generative retrieval \cite{Rajput2023RecommenderSW} has necessitated the design of novel Transformer decoding strategies. Our work lies at the intersection of large-scale generative recommendation \cite{Rajput2023RecommenderSW}, constrained token generation, and hardware-aware system optimization.

\subsection{Generative Retrieval and Semantic Indexing}
Generative retrieval (GR) represents a paradigm shift wherein the item corpus is internalized within the model parameters, replacing the traditional large-scale embedding and nearest neighbor search pipeline with a direct sequence generation task \cite{Tay2022TransformerMA}. 

Recent approaches in this area, specifically TIGER \cite{Rajput2023RecommenderSW}, SEATER \cite{Si2023GenerativeRW}, and LIGER \cite{Yang2024UnifyingGA}, utilize Semantic IDs---hierarchical, discrete token-based item representations, where semantically similar items are optimized to share the same prefix.
These generative retrieval techniques produce improved results and have been deployed in multiple online recommendation platforms, as seen in PLUM \cite{He2025PLUMAP} and OneRec \cite{Deng2025OneRecUR, Zhou2025OneRecTR, Zhou2025OneRecV2TR, Liu2025OneRecThinkIR}. However, a critical limitation of these autoregressive retrieval models is the ``validity gap.'' Unlike natural language generation, where the output space is open-ended, retrieval requires the generated string to map perfectly to a valid item index. As noted by \citet{Rajput2023RecommenderSW}, without constraints, the model effectively ``hallucinates'' non-existent inventory, leading to retrieval failures. This issue is especially pertinent in industry settings where practitioners deploy the same pre-trained model across multiple product environments, each having its own valid corpus, with some requiring candidates to fall within a certain regional locality or meet certain ``freshness'' requirements. While previous work focuses on improving recall metrics \cite{He2025PLUMAP, Si2023GenerativeRW, Zhou2025OneRecTR, Deng2025OneRecUR, Ju2025GenerativeRW}, it largely ignores the latency costs of enforcing validity constraints in a high-throughput production environment; we explicitly address this in our paper.

Additionally, existing generative retrieval papers \cite{Yang2024UnifyingGA} have noted that generative retrieval models struggle with cold-start item recommendation: the setting in which a model must recommend items it has never before seen during training. We demonstrate in Section \ref{sec:cold_start_amazon} that by implementing constrained decoding with a cold-start item set, one can improve cold-start performance significantly.

\subsection{Constrained Decoding in NLP}
Constrained decoding (CD) has a rich history in NLP.
For example, NeuroLogic \cite{Lu2021NeuroLogicAD} uses search-based heuristics to satisfy complex logical predicates, and Synchromesh \cite{Poesia2022SynchromeshRC} enforces Context-Free Grammars (CFGs) to ensure syntactic correctness in code generation. %
An approach related to our use case is that of Finite State Transducers (FSTs), which have long been the standard for constrained speech recognition \cite{Koo2024AutomatabasedCF}. However, FSTs suffer from state explosion when applied to the unstructured, high-cardinality vocabularies found in recommendation systems (often $10^7+$ items). As observed by \citet{Koo2024AutomatabasedCF}, while FSTs are powerful, their irregularity makes them difficult to parallelize on TPUs/GPUs compared to dense matrix operation-based methods.

\subsection{Hardware-Aware Inference and Acceleration}
The ``Memory Wall'' remains the primary bottleneck for Large Language Model inference. Because autoregressive decoding is memory-bandwidth bound, performance is dictated by how fast data moves to the chip's arithmetic units \cite{Dao2022FlashAttentionFA}.

\textbf{Pointer Chasing vs. Coalesced Access.} Standard trie implementations used in various constrained decoding approaches rely on pointer chasing \cite{Liao2025EliminatingOR}. On modern accelerators (TPUs/GPUs), pointer chasing induces random memory access patterns that lead to uncoalesced loads and cache thrashing. This is antithetical to the design of accelerators, which thrive on static, contiguous memory access (as seen in optimizations like FlashAttention \cite{Dao2022FlashAttentionFA}).

\textbf{Accelerator-Compatible Decoding.} The recent paper DISC-PPV \cite{Ye2025EfficientAA} represents the state-of-the-art in adapting constrained decoding to hardware accelerators. The authors implement an on-chip validity check using a flattened sorted array and parallelized binary search, a method they call Parallel Prefix-Verification (PPV). While PPV eliminates the CPU-GPU communication overhead, its reliance on binary search introduces an I/O complexity scaling factor of $O(\log |\mathcal{C}|)$ with respect to the constrained item set size $|\mathcal{C}|$. In ultra-large vocabularies (e.g. $20$ million items), even logarithmic scaling becomes an I/O bottleneck. Our introduced STATIC approach advances this line of research by replacing the $O(\log |\mathcal{C}|)$ dependent block transfers of PPV with $O(1)$ vectorized sparse matrix operations, fetching all required working sets in a single coalesced phase. 

\subsection{Linearized Tree Structures}
The concept of flattening tree structures into array-based representations dates back to the Double-Array Trie \cite{Aoe1989AnED}, originally designed to improve memory efficiency for string matching. Similarly, the GraphBLAS standard \cite{Kepner2016MathematicalFO} posits that graph algorithms can be expressed as linear algebra operations over sparse semirings.

While these concepts are established in theoretical computer science, their application to LLM constrained decoding is novel. We bridge the gap between these classical data structures and modern deep learning compilers (XLA/Inductor). By reformulating trie traversal via sparse matrix operations, we enable the use of high-performance, ML-compiler-accelerated kernels that maintain the parallelism required for real-time recommendation at scale.

\begin{figure*}[t]
  \centering
  \includegraphics[width=\linewidth]{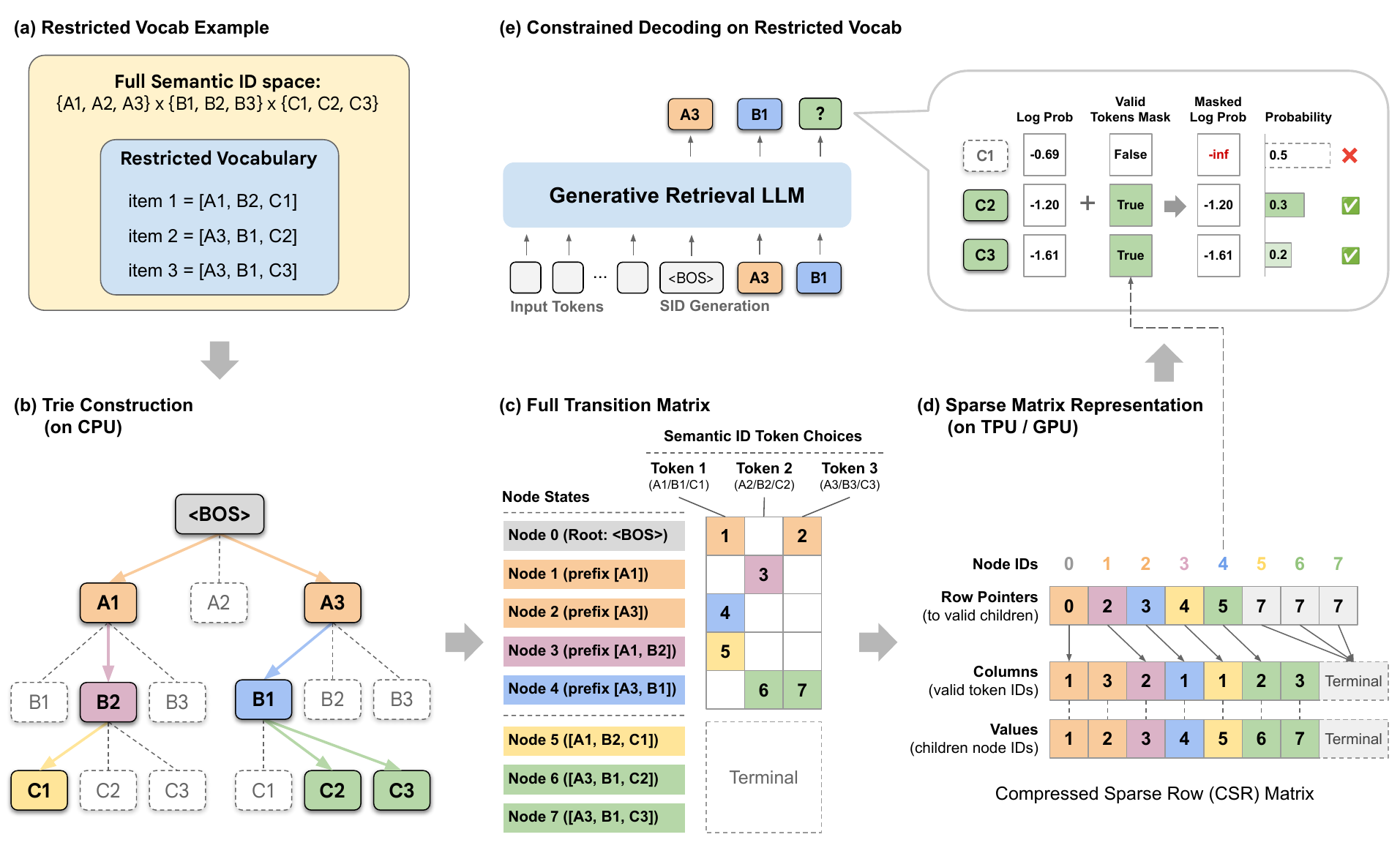}
  \caption{This figure showcases the full STATIC pipeline. Figures 1a and 1b present the prefix tree construction for the case $\mathcal{V} = \{1,2,3\}$, $L = 3$, and restricted vocabulary $\mathcal{C} = \{(1, 2, 1), (3, 1, 2), (3, 1, 3)\}$. We prepend the letters $A, B, C$ to the semantic tokens to denote first, second, and third levels, respectively. The corresponding transition matrix is then shown in Figure 1c, with explicit labels and color-coding. Figure 1d presents the sparse matrix (CSR) representation of the full transition matrix. Figure 1e shows how the sparse matrix is applied to constrain the vocabulary at decoding time.}
  \Description{Main figure for Sparse Transition Matrix-based Trie Decoding.}
  \label{fig:main_figure}
\end{figure*}

\section{Background}

In this section, we provide the background necessary to understand our paper and define the technical terminology central to generative retrieval and constrained decoding.

\subsection{Generative Retrieval and Semantic IDs}
Generative retrieval is a recommendation paradigm wherein a model directly predicts target candidate identifiers token-by-token rather than through nearest-neighbor search in an embedding space. Central to this approach is the concept of the Semantic ID: a discrete, token-based identifier that captures the semantic properties of an item. 

In the TIGER framework \cite{Rajput2023RecommenderSW}, Semantic IDs are created using a Residual-Quantized Variational AutoEncoder (RQ-VAE). An RQ-VAE first encodes item features into a latent vector $\mathbf{z}$,
which is then iteratively quantized across $L$ levels. The first-level residual is defined as: $\mathbf{r}_1 := \mathbf{z}$. At each level $d$, the residual $\mathbf{r}_{d}$ is identified with the nearest embedding $\mathbf{e}_{y_{d}}$ in a level-specific codebook $\mathcal{E}_{d}$. The index $y_{d}$ becomes the $d$-th codeword in the Semantic ID. The next-level residual is computed as: $\mathbf{r}_{d+1} := \mathbf{r}_d - \mathbf{e}_{y_d}$.

This recursive approach allows semantically similar items to share prefix tokens, enabling the model to effectively generalize across the item space. The final resulting Semantic ID is simply the tuple comprised of the embedding indices, i.e. $(y_1, \ldots, y_L)$.

\subsection{Autoregressive Decoding via Beam Search}
During inference, the retrieval model autoregressively decodes the token sequences of Semantic IDs. To generate target items we typically employ beam search~\cite{Freitag2017BeamSS} over the Semantic ID space (although our approach is also compatible with other decoding methods). 
Let $\mathcal{V}$ be the vocabulary of semantic tokens and let $L$ be the fixed length of a Semantic ID. At each step $t$, the algorithm maintains a beam search state $\mathbf{S}_t$ which tracks the $M$ most probable sequences across a batch of size $B$. We formally define $\mathbf{S}_t$ as a structure with the following members and accessors:
\begin{itemize}
    \item $\mathbf{S}_t.\text{tokens} \in \mathcal{V}^{B \times M \times t}$: The token buffer containing the decoded prefixes for each beam.
    \item $\mathbf{S}_t.\text{scores} \in \mathbb{R}^{B \times M}$: The cumulative log-probability scores for the sequences.
\end{itemize}
At step $t$, the model generates logits $\mathbf{L}_t \in \mathbb{R}^{B \times M \times |\mathcal{V}|}$ which are normalized into log-probabilities before selecting the next set of candidates. The cumulative log-probability score is updated by adding the model's predicted log-probabilities for the next token to the current prefix scores. The top $M$ candidates with the highest cumulative scores are retained, while others are pruned.

\subsection{Prefix Trees and Constrained Decoding}
Standard generative models can ``hallucinate'' identifiers that either (1) do not map to valid items in the corpus or (2) do not map to desirable items (e.g. too old, in the context of freshness). To prevent this, constrained decoding restricts the output to a restricted vocabulary set $\mathcal{C}$ containing only desirable Semantic IDs. 

Because Semantic IDs are structured as sequences, the restricted vocabulary inherently forms a prefix tree (i.e. trie). Constraints are enforced at each step $t$, that is, if appending the current token to the prefix results in a invalid path in the trie, the path is pruned. In practice, the log-probabilities of invalid tokens are set to $-\infty$, ensuring that beam search only selects valid sequences. The challenge addressed in this work is that naive pointer-chasing implementations of trie constraints incur severe latency penalties on hardware accelerators like TPUs/GPUs, making them infeasible in practice.

\section{Methodology}
\label{sec:methodology}

We propose an accelerator-native constrained decoding technique designed for LLM-based generative retrieval. Our approach replaces dynamic pointer-chasing CPU-based prefix tree traversals with a static sparse matrix lookup, enabling purely on-device execution (TPU/GPU) compatible with XLA/Inductor compilation.

\subsection{Problem Formulation}
Recall that $\mathcal{V}$ represents the set of semantic tokens (i.e. indices) in our context, and $L$, the fixed length of a Semantic ID. Note that the space of all Semantic IDs is simply the product space $\mathcal{V}^L$. Although the vocabulary of indices is the same across all levels, the codewords are different, with a different codebook $\mathcal{E}_d$ trained for each level.

The retrieval model generates a sequence (i.e. a Semantic ID) $\mathbf{y} = (y_1, \dots, y_L)$ auto-regressively, where $y_i \in \mathcal{V}$. In standard beam search, the validity of a sequence is determined solely by the model's learned probability distribution $P_\theta$. However, in constrained decoding we require the generated output to belong to a strictly defined subset $\mathcal{C} \subset \mathcal{V}^L$ (e.g. a ``Freshness'' corpus or ``Inventory'' list). An example of a restricted subset is given in Figure \ref{fig:main_figure}a.

We formally define constraint functions $F_t : \mathcal{V}^{t-1} \times \mathcal{V} \rightarrow \{0,1\}$ for $1 \leq t < L$ by:
$F_t(y_{<t}, y_t) = \mathbb{I}(\exists \:c \in \mathcal{C}\:\text{s.t.}\:(y_{<t}, y_t) \sqsubseteq c)$,
where $\mathbb{I}(\cdot)$ is the indicator function and $\sqsubseteq$ denotes prefix inclusion. $F_t(y_{<t}, y_t)$ is an indicator function that returns $1$ if and only if appending $y_t$ to the prefix $y_{<t}$ results in a valid prefix in $\mathcal{C}$. Our goal is to enforce this constraint only during inference, such that $P(y_t | y_{<t}) = 0$ if $F_t(y_{<t}, y_t) = 0$, with minimal latency overhead.

\subsection{Sparse Transition Matrix (STM)-based Trie Conversion}
Standard implementations of prefix trees rely on pointer-chasing operations that have a huge CPU callback bottleneck and are poorly supported on accelerators. To resolve this, we flatten the prefix tree into a Compressed Sparse Row (CSR) transition matrix.

Let the number of total prefix nodes in the trie be $S$. We map every unique prefix node in the tree to a state integer $s \in [S]$. We then define a sparse matrix $\mathbf{T} \in \mathbb{Z}^{S \times |\mathcal{V}|}$ where:
\[
\mathbf{T}_{s, v} =
    \begin{cases}
    s_{\text{next}} & s \xrightarrow{v} s_{next}\:\text{exists} \\
    0 & \text{otherwise} \\
    \end{cases}
\]
Conceptually, $0$ denotes a terminal state where no more transitions are possible (think of it as a ``sink''). To better illustrate the idea, we give the conceptual transition matrix for the restricted vocabulary from Figure \ref{fig:main_figure}a in Figure \ref{fig:main_figure}c (corresponding to the trie in Figure \ref{fig:main_figure}b).

\subsubsection{Matrix Construction}
The construction of $\mathbf{T}$ is performed offline. Given the high sparsity of the vocabulary constraints (often $\leq \! 0.01\%$ of valid paths), the CSR representation is highly memory efficient. In our case, we fill the three parts of the CSR representation as follows:
\begin{itemize}
    \item \textbf{Row Pointers ($\mathbf{P}$):} Stores indices for the allowed transitions for current state $s$, i.e. these are effectively node pointers.
    \item \textbf{Column Indices ($\mathbf{C}$):} Stores the valid token IDs ($v$) that trigger transitions.
    \item \textbf{Values Array ($\mathbf{V}$):} Stores the target state node IDs ($s_{next}$).
\end{itemize}
This static structure allows us to utilize hardware-optimized sparse matrix operations rather than dynamic control flow. To illustrate, for the transition matrix in Figure \ref{fig:main_figure}c we provide the CSR arrays in Figure \ref{fig:main_figure}d. Notice, for example, to extract the transitions from Node 4 in Figure \ref{fig:main_figure}c, we would first extract $\texttt{row\_start} = \textbf{P}[4] = 5$ and $\texttt{row\_end} = \textbf{P}[4+1] = 7$. Then we form the slices:
\begin{align*}
\textbf{C}[\texttt{row\_start}\!:\!\texttt{row\_end}] = \textbf{C}[5\!:\!7] = [2,3]
\\
\textbf{V}[\texttt{row\_start}\!:\!\texttt{row\_end}] = \textbf{V}[5\!:\!7] = [6,7]
\end{align*}
to observe that taking ``Token 2'' from Node 4 leads to Node 6, and taking ``Token 3'' from Node 4 leads to Node 7. One can confirm that this is indeed correct by noting the transitions present in Figure~\ref{fig:main_figure}c.

\begin{algorithm}[b]
\small
\SetAlgoLined
\DontPrintSemicolon

\SetKwFunction{LogSoftmax}{\textsc{LogSoftmax}}
\SetKwFunction{DenseLookup}{\textsc{DenseLookup}}
\SetKwFunction{VNTK}{\textsc{VNTK}}
\SetKwFunction{Where}{\textsc{Where}}
\SetKwFunction{BeamSearch}{\textsc{BeamSearch}}
\SetKwFunction{Gather}{\textsc{Gather}}
\SetKwData{Inf}{$-\infty$}

\KwIn{Logits $\mathbf{L}_t \in \mathbb{R}^{B \times M \times |\mathcal{V}|}$, Beam State $\mathbf{S}_{t-1}$, Step $t$}
\KwIn{Dense Count~$d$, Dense Restrict Tensor \scriptsize$\mathbf{D} \in \mathbb{R}^{\overbrace{|\mathcal{V}| \times \cdots \times |\mathcal{V}|}^{d\:\text{times}}}$}
\KwIn{Constraints: Trans. Matrix $\mathbf{T}$, Node Indices $\mathbf{n}_{t-1} \in \mathbb{Z}^{B \times M}$}
\KwOut{Updated State $\mathbf{S}_t$, New Node Indices $\mathbf{n}_t$}
\vspace{0.1cm}
\hrule\vspace{0.1cm}

\textbf{Phase 1: Log-Space Projection} \;
$\mathbf{P}_t \leftarrow \LogSoftmax(\mathbf{L}_t)$ \tcp*{Convert logits to log-probs}

\vspace{0.2cm}\textbf{Phase 2: Constraint Masking} \;
\tcp{For earlier layers, we employ dense lookups}
\eIf{$t-1 < d$}{ 
  $(\mathbf{n}_{next}, \mathbf{m}) \leftarrow \DenseLookup(\mathbf{n}_{t-1}, \mathbf{D}, t-1)$ \;
}{
  $(\mathbf{n}_{next}, \mathbf{m}) \leftarrow \VNTK(\mathbf{n}_{t-1}, \mathbf{T}, t-1)$ \tcp*{Execute Algorithm \ref{alg:jax_lookup}}
}
$\mathbf{P}'_t \leftarrow \Where(\mathbf{m}, \mathbf{P}_t, \Inf)$ \tcp*{Mask invalid transitions}
\vspace{0.2cm}
\textbf{Phase 3: Beam Search Optimization} \;
$(\mathbf{S}_{best}, \mathbf{I}_{best}) \leftarrow \BeamSearch(\mathbf{P}'_t, \mathbf{S}_{t-1}.\text{scores}, M)$ \;

\vspace{0.2cm}
\textbf{Phase 4: State Update} \;
$\mathbf{S}_t.\text{tokens} \leftarrow \Gather(\mathbf{S}_{t-1}.\text{tokens}, \text{indices}=\mathbf{I}_{best})$ \;
$\mathbf{S}_t.\text{scores} \leftarrow \mathbf{S}_{best}$ \;
$\mathbf{n}_t \leftarrow \Gather(\mathbf{n}_{next}, \text{indices}=\mathbf{I}_{best})$ \tcp*{Advance constraint state}

\vspace{0.1cm}
\Return $\mathbf{S}_t, \mathbf{n}_t$ \;
\caption{Hardware-Accelerated Constrained Decoding Step}
\label{alg:constrained_decoding}
\end{algorithm}

\subsection{Accelerator-Native Decoding Algorithm}
To achieve high throughput on hardware accelerators, we reformulate the validity check as a vectorized lookup. Critically, we maintain a transition state vector $\mathbf{n}_t \in \mathbb{Z}^{B \times M}$ that tracks the current node index in the prefix tree for each beam. Additionally, we employ the following optimization. For a number of layers $0 \leq d < L$, we maintain a dense tensor mask $\mathbf{D} \in \mathbb{R}^{\overbrace{|\mathcal{V}| \times \cdots \times |\mathcal{V}|}^{d\:\text{times}}}$ that indicates precisely which $d$-length prefixes exist in the restricted vocabulary $\mathcal{C}$. Lookups in this tensor are extremely fast and the construction of $\mathbf{D}$ is a one-time fixed cost; however, $\mathbf{D}$ becomes prohibitively expensive to construct for $d \geq 3$ in most cases due to the fact that $|\mathcal{V}|^d$ grows exponentially in $d$. Hence, we have $d \leq 2$ for most real-world use cases and rely on the sparse CSR transition matrix for deeper layers. The full sequential constrained decoding process is given in Algorithm \ref{alg:constrained_decoding}. Our approach relies on several hardware-accelerated operations:
\begin{itemize}
    \item $\textbf{LogSoftmax}(\mathbf{x})$: Converts raw logits into normalized log-probabilities according to the following transformation:
    $
    \text{LogSoftmax}(\mathbf{x})_i = x_i - \log \sum_{j=1}^{|\mathcal{V}|} \exp(x_j)
    $.

    \item $\textbf{DenseLookup}(\mathbf{n}_{t-1}, \mathbf{D}, t-1)$: This primitive performs a high-performance lookup on the dense tensor mask $\mathbf{D}$ to identify valid semantic extensions for each beam. 
    
    \item $\textbf{VNTK}(\mathbf{n}_{t-1}, \mathbf{T}, t-1)$: The \textit{Vectorized Node Transition Kernel} (Algorithm \ref{alg:jax_lookup}). It performs a high-performance lookup on the sparse transition matrix $\mathbf{T}$ to identify valid semantic extensions for each beam. It encapsulates the speculative slicing and projection logic required to return the next-node IDs $\mathbf{n}_{next}$ and the dense log probability mask $\mathbf{m}$ in a single vectorized pass. See Section \ref{sec:implementation} for more detail.
    
    \item $\textbf{Where}(\text{mask}, \mathbf{x}, \text{value})$: This conditional operator enforces validity constraints in log-probability space. It prunes invalid paths by setting their scores to $-\infty$ where the mask $\mathbf{m}$ is false.
    
    \item $\textbf{BeamSearch}(\mathbf{P}'_t, \mathbf{S}_{scores}, M)$: A selection framework that identifies $M$ most probable valid sequences across the batch. It computes cumulative scores by combining the new masked log-probabilities $\mathbf{P}'_t$ with existing prefix scores $\mathbf{S}_{scores}$ and retains the top candidates for the next step.
    
    \item $\textbf{Gather}(\mathbf{x}, \text{indices})$: A vectorized selection operator used in the state update phase (Phase 4). It extracts the specific tokens, scores, and constraint states from the candidate pool that correspond to the top-performing beams ($\mathbf{I}_{best}$).
\end{itemize}

\subsection{Implementation and Hardware Optimization}
\label{sec:implementation}

While the theoretical formulation of STATIC guarantees correctness, achieving negligible latency overhead on modern accelerators requires rigorous alignment with the underlying hardware characteristics. In this section, we detail the vectorized kernel design that makes our approach portable across both TPUs and GPUs. 

\begin{algorithm}[b]
\small
\SetAlgoLined
\DontPrintSemicolon

\SetKwFunction{DSlice}{\textsc{DynamicSlice}}
\SetKwFunction{Range}{\textsc{Range}}
\SetKwFunction{Where}{\textsc{Where}}
\SetKwFunction{Scatter}{\textsc{Scatter}}
\SetKwData{PadTok}{$\{\}$}
\SetKwData{PadNode}{$0$}

\KwIn{Curr. Node $n_{curr}$, Trans. Matrix $\mathbf{T}$ (Row Pointers $\mathbf{P}$, Columns $\mathbf{C}$, Values $\mathbf{V}$), Step $t$}
\KwIn{(Constant) Token Cardinality $|\mathcal{V}|$, Max Branch Factors $\mathbf{B}$}
\KwOut{Next Nodes $\mathbf{n}_{next}$, Logit Mask $\mathbf{m}$}
\vspace{0.1cm}
\hrule\vspace{0.1cm}

\textbf{Phase 1: Boundary Lookup} \;
$idx_{start} \leftarrow \mathbf{P}[n_{curr}]$ \;
$N_{child} \leftarrow \mathbf{P}[n_{curr} + 1] - idx_{start}$ \tcp*{Compute \# of children}

\vspace{0.2cm}
\textbf{Phase 2: Speculative Slicing} \;
\tcp{Slice fixed size $\mathbf{B}_t$ regardless of actual child count}
$\mathbf{d}_{col} \leftarrow \DSlice(\mathbf{C}, \text{start}=idx_{start}, \text{len}=\mathbf{B}_t)$ \;
$\mathbf{d}_{val} \leftarrow \DSlice(\mathbf{V}, \text{start}=idx_{start}, \text{len}=\mathbf{B}_t)$ \;

\vspace{0.2cm}
\textbf{Phase 3: Sanitization (Branch-Free)} \;
$J \leftarrow \Range(\mathbf{B}_t)$ \tcp*{Generate vector $[0, \dots, \mathbf{B}_t-1]$}
$\mathbf{m}_{valid} \leftarrow (J < N_{child})$ \tcp*{Identify valid slots}

\vspace{0.1cm}
$\mathbf{t}_{valid} \leftarrow \Where(\mathbf{m}_{valid}, \mathbf{d}_{col}, \PadTok)$ \;
$\mathbf{n}_{next} \leftarrow \Where(\mathbf{m}_{valid}, \mathbf{d}_{val}, \PadNode)$ \;

\vspace{0.2cm}
\textbf{Phase 4: Projection} \;
\tcp{Create dense boolean mask for Softmax}
$\mathbf{m} \leftarrow \Scatter(\text{indices}=\mathbf{t}_{valid}, \text{values}=\mathbf{m}_{valid})$ \;

\vspace{0.1cm}
\Return $\mathbf{n}_{next}, \mathbf{m}$ \;
\caption{Vectorized Node Transition Kernel (VNTK)}
\label{alg:jax_lookup}
\end{algorithm}

The core engineering challenge in constrained decoding is handling the dynamic nature of the prefix tree (where nodes possess a varying number of children) within the static execution model of hardware accelerators. Standard pointer-chasing traversals are inefficient: on TPUs (XLA), dynamic branching triggers costly graph recompilation or forces the use of serial `while' loops that prevent hardware pipelining; on GPUs, processing beams with mismatched child counts leads to GPU warp divergence, where threads must wait for the longest path to finish, destroying SIMT parallelism.

To overcome these bottlenecks, we implement a branch-free transition kernel (Algorithm \ref{alg:jax_lookup}). For efficient slicing, it becomes critical to introduce some new definitions. Consider the prefix tree corresponding to a sparse transition matrix $\mathbf{T}$. For a node $n$ in the tree at level $\ell$, let the number of children be $b^\ell_n$. Given that the tree has $L$ levels, let the set of all nodes at level $\ell \in [L]$ be $N_\ell$ and define the vector $\mathbf{B} \in \mathbb{R}^L$ such that $\mathbf{B}_\ell=\max_{n \in N_\ell} b^\ell_n$. That is, $\mathbf{B}$ is the level-indexed vector of max branch factors exhibited by the tree. Note that computing this vector is a one-time fixed cost per transition matrix $\mathbf{T}$. Our algorithm takes this vector as an input. Moreover, Algorithm \ref{alg:jax_lookup} relies on the following vectorized primitive operations:
\begin{itemize}
    \item $\textbf{DynamicSlice}(\mathbf{M}, \text{start}, \text{len})$: Extracts a contiguous sub-tensor of fixed length $len = \mathbf{B}_t$ from the data matrix $\mathbf{M}$ starting at a dynamic offset ($start$). Recall decoding step $t$ is synonymous with level $\ell$ in the underlying prefix tree. 
    
    \item $\textbf{Range}(c)$: Generates a constant sequence vector $[0, \dots, c-1]$ used for index-based masking and sanitization of the speculatively sliced children.

    \item $\textbf{Scatter}(\text{indices}, \text{values})$: A projection operator that converts the sparse list of valid tokens into a dense boolean mask $\mathbf{m}$ of size $|\mathcal{V}|$. This mask is applied directly to the model's log-probabilities in the main decoding step to enforce the prefix-tree constraints.
\end{itemize}

As shown in Phase 2, the kernel always slices a fixed number of entries $\mathbf{B}_t$ for any given level $t$ regardless of the actual child count. If a node has fewer than $\mathbf{B}_t$ children, we use a \textbf{Range}-generated vector to create a boolean mask $\mathbf{m}_{valid}$ and the \textbf{Where} operator to sanitize the result. This ensures that the arithmetic units of the GPU/TPU remain fully saturated and that the entire decoding step remains a single, static computation graph. We discuss additional relevant hardware considerations of our method in Appendix \ref{app:hardware_detailed}. We also benchmark our method for high maximum branch factors in Appendix \ref{app:appendix_branch_scaling} and demonstrate linear scaling on this axis. Importantly, we note that for sufficiently large $|\mathcal{V}|$ and practical ranges of $|\mathcal{C}|$, the max branch factor will actually remain quite low for later layers, since the number of children $|\mathcal{V}|^\ell$ grows exponentially with level $\ell$ and quickly exceeds the constraint set size $|\mathcal{C}|$.

\section{Large-scale Deployment on YouTube}
\label{sec:experiment_yt}

By testing STATIC at YouTube scale, we seek to evaluate the STATIC algorithm along three dimensions critical for industrial deployment:

\begin{enumerate}
    \item \textbf{System Efficiency}: We measure latency and throughput on TPU accelerators compared to baselines.
    \item \textbf{Scalability}: We analyze the storage footprint and runtime as both constraint set size and vocabulary size increase.
    \item \textbf{Online Product Impact}: We quantify the improvement of recommendation quality and user satisfaction from applying constrained decoding in a live production environment.
\end{enumerate}

\subsection{Experimental Setup}

We conduct our evaluation on a large-scale video recommendation corpus from YouTube. The videos are tokenized into Semantic IDs (SIDs) \cite{Rajput2023RecommenderSW} with $L=8$ discrete tokens and token cardinality $|\mathcal{V}| = 2048$. The total recommendable corpus is on the order of $100$ million to $1$ billion items, and the unconstrained model can recommend any item from this set. We fix $d=2$ as an input to Algorithm \ref{alg:constrained_decoding} and maintain the dense tensor mask $\mathbf{D} \in \mathbb{R}^{2048 \times 2048}$ for early lookups; we use the CSR sparse matrix representation for later levels.

The model is a Gemini-based generative retrieval model similar to PLUM \cite{He2025PLUMAP}, served with a batch size of 2 (per chip) and a beam size of $M=70$. The model is based on a non-Mixture-of-Experts (MoE) architecture with 3 billion dense parameters. All benchmark experiments are conducted on Google TPU v6e accelerators.

\begin{table}[b]
\centering
\caption{Latency Overhead per Decoding Step on a Large-scale Video Recommendation Corpus with $\mathbf{20}$ Million Constrained Vocabulary Items. STATIC (Ours) achieves minimal latency overhead relative to baselines.}
\label{tab:latency_bench_yt}
\resizebox{\columnwidth}{!}{%
\begin{tabular}{lcc} 
\toprule
\textbf{Method} & \textbf{Latency (ms)} & \textbf{\% of Inference} \\ 
\midrule
Unconstrained & $+0.0$ & -- \\
\midrule
PPV Exact \cite{Ye2025EfficientAA} & $+34.1$  & $260\%$ \\
CPU Trie & $+31.3$ & $239\%$ \\
Hash Bitmap & $+12.3$ & $94.0\%$ \\
PPV Approximate \cite{Ye2025EfficientAA} & $+1.56$  & $11.9\%$ \\
\textbf{STATIC (Ours)} & $\mathbf{+0.033}$ & $\mathbf{0.25\%}$\\
\bottomrule
\end{tabular}
}
\end{table}

\subsection{System Efficiency Analysis}
\label{sec:yt_system_efficiency_analysis}

We limit our attention to a constrained vocabulary (target set) consisting of a ``fresh video'' subset containing approximately $20$ million high-quality items uploaded in the last $7$ days (relative to the experiment date). This setup represents a restrictive environment for constrained decoding: the valid subset is sparse, yet large enough ($20$ million items) to make naive enumeration infeasible. 

We compare our STATIC approach against several distinct implementations\footnote{We did not compare against Finite State Transducers (FSTs) for fundamental reasons we highlight in Appendix \ref{app:fst}.} commonly found in academic literature and production systems:
\begin{enumerate}
    \item \textbf{Unconstrained:} Standard beam search with no validity checks. This represents the lower bound on latency.
    \item \textbf{CPU Trie:} A standard prefix tree stored in CPU memory using JAX's PyTree data structure (Python dictionary). At every decoding step, the TPU halts, sends partial beams to the CPU, waits for the validity mask, and resumes.
    \item \textbf{PPV Exact \cite{Ye2025EfficientAA}:} Following \citet{Ye2025EfficientAA}, we store valid SIDs in a sorted flat array on TPU and perform binary search to verify every candidate extension. In the original paper, the authors consider only the top $50$ logits of the vocabulary per decoding step. This is suboptimal for beam search with larger fan-out since constrained set items frequently fall outside of the top $50$ logits. Thus, in this baseline, we consider the PPV method using all $2048$ logits. This yields an exact solution but is significantly slower than PPV Approximate below, due to PPV's linear scaling with the number of logits.
    \item \textbf{PPV Approximate \cite{Ye2025EfficientAA}:} Just as above, we perform binary search on a sorted flat SID array for every candidate extension; however, as in the original paper, this baseline only verifies the top $50$ logits per decoding step. This gives only approximate solutions for the reasons mentioned above, and is not directly comparable to STATIC.
    \item \textbf{Hash Bitmap:} A Bloom-filter \cite{Bloom1970SpacetimeTI} style approach where valid prefixes are hashed. This method introduces false positives (with a non-negligible 2.1\% false positive rate in our Table \ref{tab:latency_bench_yt} evaluation), so it is not directly comparable to STATIC, but useful as a reference.
\end{enumerate}

STATIC is a lossless implementation optimization of the trie-based constrained decoding algorithm, so we focus on measuring the latency of the constraint enforcement logic per decoding step. Our results are summarized in Table \ref{tab:latency_bench_yt}, where we give means over $100$ trials. Our STATIC approach achieves a latency overhead of +$0.033$ ms per decoding step, only taking $0.25$\% of the inference time for a 3 billion parameter model. In contrast, the ``CPU Trie'' method incurs a massive penalty ($31.3$ ms, $948\text{\small $\times$}$ higher than STATIC) due to the PCIe transfer overhead and synchronization locks between the TPU and CPU. Crucially, we outperform the ``PPV Exact'' baseline by a factor of $1033\text{\small $\times$}$ ($34.1$ ms vs. $0.033$ ms) and even the ``PPV Approximate'' baseline by factor of $47\text{\small $\times$}$ ($1.56$ ms vs. $0.033$ ms). While PPV's binary search is technically ``on-device'' and vectorized, it suffers from logarithmic scaling of I/O complexity with respect to the constraint set size $|\mathcal{C}|$ (on top of linear scaling with respect to number of logits considered), whereas STATIC reduces the I/O complexity to $O(1)$ using the single-pass VNTK kernel (Algorithm~\ref{alg:jax_lookup}).

\subsection{Scalability}

While a naive boolean mask of the entire Semantic ID space would be petabytes in size, our STATIC approach requires only approximately $90$ MB of memory for every 1 million restricted vocabulary items in this setting. That is, for our tested restricted vocabulary of $20$ million items, we expect a maximal HBM usage of $\approx 1.8$ GB, which we can further tighten to $\approx 1.5$ GB via the analysis in Appendix \ref{app:static_memory_analysis}. In practice, the required memory is usually $\leq 75\%$ of this upper limit. We elaborate on the derivations and why this is the case in Appendix \ref{app:static_memory_analysis}. The crucial takeaway is that the memory usage is tied directly to the storage of the dense tensor mask $\mathbf{D}$ and the sparse CSR matrix, which is tied to the number of nodes in the underlying trie. Given a vocabulary size $|\mathcal{V}|$, Semantic ID length $L$, dense layer count $d$, number of constraints $|\mathcal{C}|$, storage cost per node of $K_1$ bytes, and storage cost per dense state of $K_2$ bytes, the upper bound on memory usage (in bytes) is given by:
\[
U_{\text{max}}(K_1, K_2, |\mathcal{V}|, |\mathcal{C}|, L, d) = \left(\frac{1}{8} +K_2\right) |\mathcal{V}|^d + K_1 \cdot \sum_{\ell=d+1}^{L} \min(|\mathcal{V}|^\ell, |\mathcal{C}|)
\]

\begin{figure}[t]
  \centering
  \includegraphics[width=\linewidth]{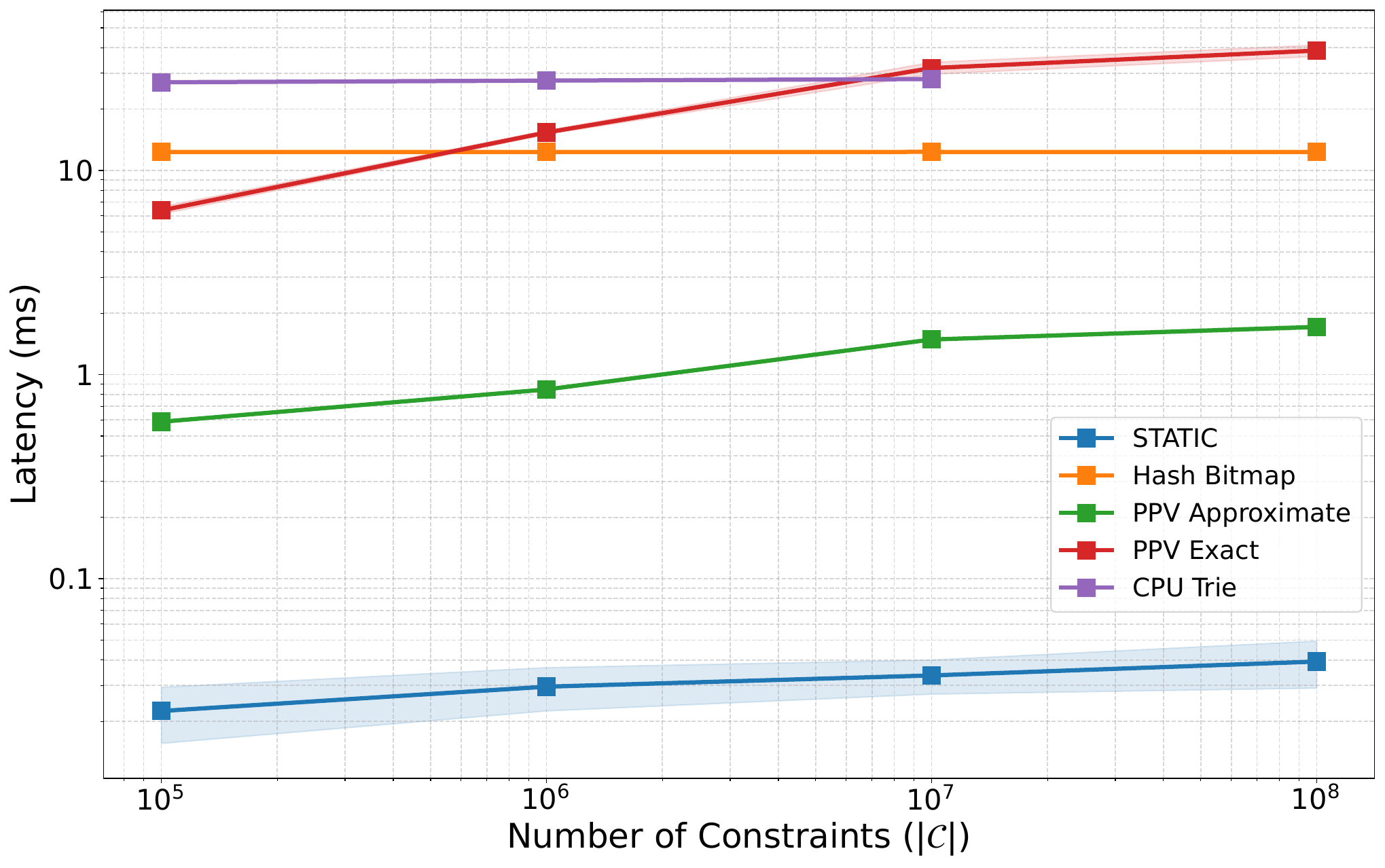}
  \caption{Scaling of different constraint decoding methods relative to constraint set size (log-log scale), fixing $|\mathcal{V}|=2048$. STATIC considerably outperforms existing methods.}
  \label{fig:scaling_constraints}
\end{figure}

Additionally, STATIC's time complexity scales logarithmically with respect to constraint set size and maintains low latency relative to all measured baselines across a wide range of $|\mathcal{C}|$. A direct comparison with the Table \ref{tab:latency_bench_yt} baselines is given in Figure \ref{fig:scaling_constraints}, where we generate constraint sets uniformly at random, fixing $|\mathcal{V}|=2048$, $L=8$ and vary only $|\mathcal{C}|$ from $10^5$ to $10^8$. The means over $100$ trials are plotted and the shaded region indicates one standard deviation above and below the mean\footnote{The $|\mathcal{C}|=10^8$ data point for CPU Trie is missing due to an Out-Of-Memory error.}.

\begin{figure}[t]
  \centering
  \includegraphics[width=\linewidth]{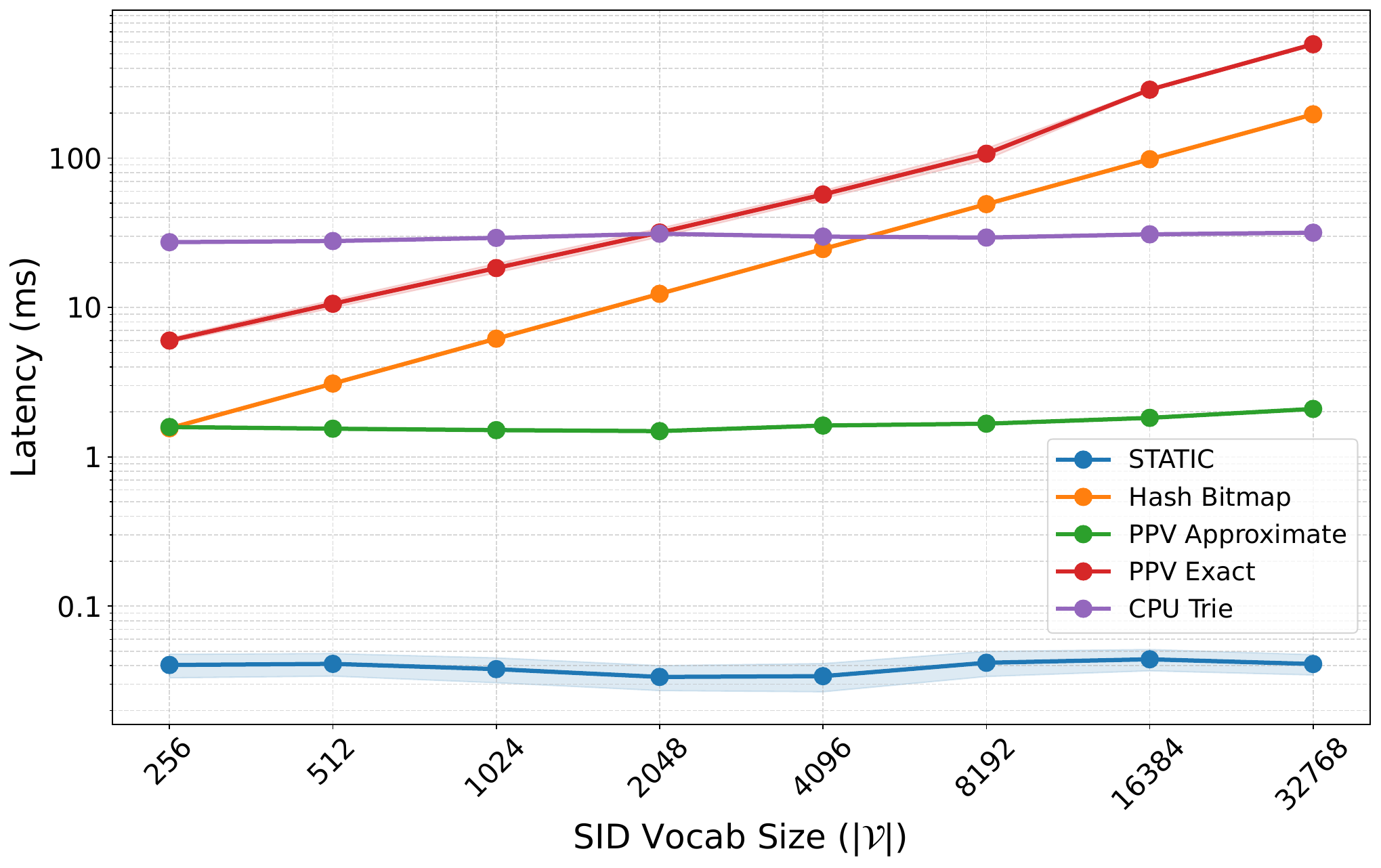}
  \caption{Scaling of different constraint decoding methods relative to SID vocabulary size $|\mathcal{V}|$ (log-log scale) at $|\mathcal{C}|=10^7$. STATIC compares favorably for all tested SID vocab sizes.}
  \label{fig:scaling_vocab_size}
\end{figure}

Furthermore, STATIC exhibits almost constant latency with respect to the SID vocabulary size $|\mathcal{V}|$, as shown in Figure \ref{fig:scaling_vocab_size}, where we generate constraint sets uniformly at random just as above, fixing $|\mathcal{C}|=10^7$, $L=8$, and vary only $|\mathcal{V}|$ from 256 to 32k. In theory, the pure sparse matrix look-ups scale with maximum branch factor (as shown in Algorithm \ref{alg:jax_lookup}), whose average (under mild assumptions) scales linearly with $|\mathcal{V}|$. However, since we employ a dense mask with $O(1)$ look-up for the highly saturated first $d=2$ layers, the sparse look-ups happen for later layers with small constant maximum branch factors. As such, STATIC exhibits near constant scaling with $|\mathcal{V}|$. In almost all production systems \cite{He2025PLUMAP, Zhou2025OneRecTR} as of $2025$, $|\mathcal{V}|\leq 8192$, meaning STATIC exhibits consistently low latency across all current real-world settings. Exact numbers for Figures \ref{fig:scaling_constraints} and \ref{fig:scaling_vocab_size} are given in Appendix \ref{app:latency_scaling}.

\subsection{Online A/B Testing}

We deployed a STATIC-enabled PLUM-based generative retrieval model across multiple use cases in the YouTube ecosystem. In one setting, we used it
as an additional candidate source to a short-form video product's ``Home Feed'' with a ``Last 7 Days'' freshness constraint,
and conducted A/B experiments to measure the impact. The results are shown in Table \ref{tab:online_results}.

As expected, STATIC achieved 100\% compliance with the specified constraint. A standard generative retrieval model without this constraint would frequently generate SIDs for older videos, whereas the STATIC model strictly produced valid 7-day fresh videos.

We observed a significant increase in the consumption of fresh content: as given in Table \ref{tab:online_results}, we note a $+5.1\%$ increase in 7-day fresh video views and a $+2.9\%$ increase in 3-day fresh video views.

Moreover, we observed a positive impact on user experience metrics: the click-through rate (CTR) increased by +$0.15\%$ and user satisfaction of a strategic user segment increased by +$0.15$\%. This is a significant improvement and shows the positive impact of recommending more high-quality fresh videos.

\begin{table}[t]
\centering
\caption{Online A/B Test Results for ``Home Feed'' Impact with the ``Last 7 Days'' Freshness Constraint.}
\label{tab:online_results}
\resizebox{\columnwidth}{!}{%
\begin{tabular}{c c c} 
\toprule
\textbf{Metric} & \textbf{Impact} & \textbf{$95\%$ Confidence Interval} \\ 
\midrule
7 Day Fresh Views & $+5.1\%$ & [$5.0\%$, $5.2\%$] \\
3 Day Fresh Views & $+2.9\%$ & [$2.8\%$, $3.0\%$] \\
Click-Through Rate & $+0.15\%$ & [$0.01\%$, $0.29\%$] \\
Strategic User Segment Satisfaction & $+0.15\%$ & [$0.03\%$, $0.27\%$] \\
\bottomrule
\end{tabular}
}
\end{table}

\section{Cold-Start Retrieval on Amazon Reviews Datasets}
\label{sec:cold_start_amazon}

A long-standing drawback of generative retrieval is its inability to generalize to cold-start items \cite{Yang2024UnifyingGA}. Here, we show how one can use constrained decoding on the cold-start item set to considerably improve cold-start performance. We use the Amazon Reviews \cite{He2016UpsAD} datasets to showcase a viable set-up and the associated benefits.

\subsection{Experimental Setup}

The Amazon items are tokenized into Semantic IDs (SIDs) by training separate RQ-VAE \cite{Rajput2023RecommenderSW} models to quantize the items of each Amazon Reviews subdataset separately, with $L=4$ discrete tokens and token cardinality $|\mathcal{V}| = 256$. Each subdataset's valid item corpus consists of anywhere from $10$ to $20$ thousand items, and the unconstrained model can recommend any item from this corpus.

We train a separate generative retrieval model on the sequences from each Amazon subdataset based on the Gemma \cite{Kamath2025Gemma3T} architecture with 1 billion dense parameters, served with a batch size of $16$ and a beam size of $M=20$. All experiments are conducted on a Google TPU v6e.

\subsection{Cold-Start Results}

We perform cold-start experiments in the following way. For a given Amazon item, we define its ``age'' by the age of its oldest review. Thereafter, for each Amazon subdataset, we isolate the newest $2\%$ (and separately, $5\%$) of the items into a cold-start set. We filter out all item sequences that do not contain any items from this set, and train the models on these ``no cold-start train sequences.'' We evaluate the models on the ``cold-start test sequence'' set which contains sequences that have cold-start items as targets. We tested the following approaches for retrieval:
\begin{enumerate}
    \item \textbf{Unconstrained:} Beam search with no validity checks.
    \item \textbf{Constrained Random Guessing:} A baseline that guesses items uniformly at random from the constraint set.
    \item \textbf{Dense Retrieval \cite{Hou2022TowardsUS}:} A separately trained two-tower model whose framework follows a variant of dense retrieval as presented by \citet{Hou2022TowardsUS}. We give the architectural and training details for this model in Appendix \ref{appendix:cold_start}.
    \item \textbf{STATIC (Ours):} Our proposed method, where the model decodes from the Transformer trained in (1) while restricting to the cold-start item set at every step.
\end{enumerate}

Table \ref{tab:amazon_cold} shows Recall@1 metrics for the ``cold-start test sequences.'' In both the $2\%$ and $5\%$ cold-start settings, STATIC improves over the unconstrained, constrained random guessing, and dense retrieval approaches considerably, showing that one can attain nontrivial cold-start performance with generative retrieval using constrained decoding \textit{alone}. Additional context about the evaluation protocol is given in Appendix \ref{appendix:cold_start}.

\begin{table}[h]
\centering
\caption{Comparison of Unconstrained and Constrained Decoding Performance on Amazon Cold-start Items. All values given are Recall@1 percentages.}
\label{tab:amazon_cold}
\resizebox{\columnwidth}{!}{%
\begin{tabular}{lccc}
\toprule
\textbf{Subdataset} & \textbf{Method} & \textbf{$2\%$ Cold-start} & \textbf{$5\%$ Cold-start} \\
\midrule
\multirow{4}{*}{Beauty} & Unconstrained & $0.00\%$ & $0.00\%$ \\
& Const. Random & $0.42\%$ & $0.17\%$ \\
& Dense Retrieval & $0.88\%$ & $0.63\%$ \\
& \textbf{STATIC} & $\mathbf{4.29\%}$ & $\mathbf{1.60\%}$ \\
\midrule
\multirow{4}{*}{Sports \& Outdoors} & Unconstrained & $0.00\%$ & $0.00\%$ \\
& Const. Random & $0.27\%$ & $0.11\%$ \\
& Dense Retrieval & $0.85\%$ & $0.36\%$ \\
& \textbf{STATIC} & $\mathbf{1.24\%}$ & $\mathbf{1.17\%}$ \\
\midrule
\multirow{4}{*}{Toys \& Games} & Unconstrained & $0.00\%$ & $0.00\%$ \\
& Const. Random & $0.42\%$ & $0.17\%$ \\
& Dense Retrieval & $0.49\%$ & $0.44\%$ \\
& \textbf{STATIC} & $\mathbf{4.39\%}$ & $\mathbf{2.25\%}$ \\
\bottomrule
\end{tabular}
}
\end{table}

\section{Conclusion}
\label{sec:conclusion}

Generative retrieval holds the promise of leveraging LLMs for next-generation recommendations in myriad real-world settings, but its adoption in industry has been hindered by the lack of a controllable output space. In this work, we bridged the gap between prefix tree-based constraints and vector-based hardware (TPUs/GPUs). Regarding trie-constrained decoding, the authors of \citet{Ye2025EfficientAA} claim: ``whether an efficient GPU trie algorithm exists remains to be [seen].'' This is a hypothesis that we answer in the affirmative with the contribution in our paper. The STATIC framework transforms pointer-chasing trie lookups into vectorized sparse matrix operations, achieving a $47$--$1033\text{\small $\times$}$ speedup over alternative on-device methods. We demonstrated the effectiveness of this approach in production environments serving billions of users, showing that strict constraints can be enforced at scale without compromising serving latency. Moreover, we demonstrated the viability of improving cold-start performance through constrained-decoding alone on the Amazon Reviews \cite{He2016UpsAD} datasets. 

\textbf{Future Work.} Currently the sparse transition matrix construction is an offline process. An important future extension is to develop dynamic sparse updates, which would allow for real-time inventory changes without full model recompilation.


\bibliographystyle{ACM-Reference-Format}
\bibliography{constrained}

@article{Rajput2023RecommenderSW,
  title={Recommender Systems with Generative Retrieval},
  author={Shashank Rajput and Nikhil Mehta and Anima Singh and Raghunandan H. Keshavan and Trung Hieu Vu and Lukasz Heldt and Lichan Hong and Yi Tay and Vinh Q. Tran and Jonah Samost and Maciej Kula and Ed H. Chi and Maheswaran Sathiamoorthy},
  journal={ArXiv},
  year={2023},
  volume={abs/2305.05065},
  url={https://api.semanticscholar.org/CorpusID:258564854}
}

@inproceedings{Guo2019AcceleratingLI,
  title={Accelerating Large-Scale Inference with Anisotropic Vector Quantization},
  author={Ruiqi Guo and Philip Sun and Erik M. Lindgren and Quan Geng and David Simcha and Felix Chern and Sanjiv Kumar},
  booktitle={International Conference on Machine Learning},
  year={2019},
  url={https://api.semanticscholar.org/CorpusID:218614141}
}

@article{Sun2024SOARII,
  title={SOAR: Improved Indexing for Approximate Nearest Neighbor Search},
  author={Philip Sun and David Simcha and Dave Dopson and Ruiqi Guo and Sanjiv Kumar},
  journal={ArXiv},
  year={2024},
  volume={abs/2404.00774},
  url={https://api.semanticscholar.org/CorpusID:268030651}
}

@article{Ye2025EfficientAA,
  title={Efficient and Asymptotically Unbiased Constrained Decoding for Large Language Models},
  author={Haotian Ye and Himanshu Jain and Chong You and Ananda Theertha Suresh and Haowei Lin and James Zou and Felix X. Yu},
  journal={ArXiv},
  year={2025},
  volume={abs/2504.09135},
  url={https://api.semanticscholar.org/CorpusID:277781124}
}

@inproceedings{Ju2025GenerativeRW,
    author = {Ju, Clark Mingxuan and Collins, Liam and Neves, Leonardo and Kumar, Bhuvesh and Wang, Louis Yufeng and Zhao, Tong and Shah, Neil},
    title = {Generative Recommendation with Semantic IDs: A Practitioner's Handbook},
    year = {2025},
    isbn = {9798400720406},
    url = {https://doi.org/10.1145/3746252.3761612},
    doi = {10.1145/3746252.3761612},
    booktitle = {Proceedings of the 34th ACM International Conference on Information and Knowledge Management},
    pages = {6420–6425},
    numpages = {6},
    series = {CIKM '25}
}

@inproceedings{weller2026on,
    title={On the Theoretical Limitations of Embedding-Based Retrieval},
    author={Orion Weller and Michael Boratko and Iftekhar Naim and Jinhyuk Lee},
    booktitle={The Fourteenth International Conference on Learning Representations},
    year={2026},
    url={https://openreview.net/forum?id=k9CzIvzfaA}
}

@misc{zhou2026openonerectechnicalreport,
  title={OpenOneRec Technical Report}, 
  author={Guorui Zhou and Honghui Bao and Jiaming Huang and Jiaxin Deng and Jinghao Zhang and Junda She and Kuo Cai and Lejian Ren and Lu Ren and Qiang Luo and Qianqian Wang and Qigen Hu and Rongzhou Zhang and Ruiming Tang and Shiyao Wang and Wuchao Li and Xiangyu Wu and Xinchen Luo and Xingmei Wang and Yifei Hu and Yunfan Wu and Zhanyu Liu and Zhiyang Zhang and Zixing Zhang and Bo Chen and Bin Wen and Chaoyi Ma and Chengru Song and Chenglong Chu and Defu Lian and Fan Yang and Feng Jiang and Hongtao Cheng and Huanjie Wang and Kun Gai and Pengfei Zheng and Qiang Wang and Rui Huang and Siyang Mao and Tingting Gao and Wei Yuan and Yan Wang and Yang Zhou and Yi Su and Zexuan Cheng and Zhixin Ling and Ziming Li},
  year={2026},
  eprint={2512.24762},
  archivePrefix={arXiv},
  primaryClass={cs.IR},
  url={https://arxiv.org/abs/2512.24762}, 
}

@article{Koo2024AutomatabasedCF,
  title={Automata-based constraints for language model decoding},
  author={Terry Koo and Frederick Liu and Luheng He},
  journal={ArXiv},
  year={2024},
  volume={abs/2407.08103},
  url={https://api.semanticscholar.org/CorpusID:271097802}
}

@article{Si2023GenerativeRW,
  title={Generative Retrieval with Semantic Tree-Structured Identifiers and Contrastive Learning},
  author={Zihua Si and ZhongXiang Sun and Jiale Chen and Guozhang Chen and Xiaoxue Zang and Kai Zheng and Yang Song and Xiao Zhang and Jun Xu},
  journal={Proceedings of the 2024 Annual International ACM SIGIR Conference on Research and Development in Information Retrieval in the Asia Pacific Region},
  year={2023},
  url={https://api.semanticscholar.org/CorpusID:262466132}
}

@inproceedings{Lu2021NeuroLogicAD,
  title={NeuroLogic A*esque Decoding: Constrained Text Generation with Lookahead Heuristics},
  author={Ximing Lu and Sean Welleck and Peter West and Liwei Jiang and Jungo Kasai and Daniel Khashabi and Ronan Le Bras and Lianhui Qin and Youngjae Yu and Rowan Zellers and Noah A. Smith and Yejin Choi},
  booktitle={North American Chapter of the Association for Computational Linguistics},
  year={2021},
  url={https://api.semanticscholar.org/CorpusID:245218671}
}

@article{Poesia2022SynchromeshRC,
  title={Synchromesh: Reliable code generation from pre-trained language models},
  author={Gabriel Poesia and Oleksandr Polozov and Vu Le and Ashish Tiwari and Gustavo Soares and Christopher Meek and Sumit Gulwani},
  journal={ArXiv},
  year={2022},
  volume={abs/2201.11227},
  url={https://api.semanticscholar.org/CorpusID:246294475}
}

@article{Raffel2019ExploringTL,
  title={Exploring the Limits of Transfer Learning with a Unified Text-to-Text Transformer},
  author={Colin Raffel and Noam Shazeer and Adam Roberts and Katherine Lee and Sharan Narang and Michael Matena and Yanqi Zhou and Wei Li and Peter J. Liu},
  journal={J. Mach. Learn. Res.},
  year={2019},
  volume={21},
  pages={140:1-140:67},
  url={https://api.semanticscholar.org/CorpusID:204838007}
}

@inproceedings{Freitag2017BeamSS,
  title={Beam Search Strategies for Neural Machine Translation},
  author={Markus Freitag and Yaser Al-Onaizan},
  booktitle={NMT@ACL},
  year={2017},
  url={https://api.semanticscholar.org/CorpusID:2229477}
}

@article{Bloom1970SpacetimeTI,
  title={Space/time trade-offs in hash coding with allowable errors},
  author={Burton H. Bloom},
  journal={Commun. ACM},
  year={1970},
  volume={13},
  pages={422-426},
  url={https://api.semanticscholar.org/CorpusID:7931252}
}

@article{Tay2022TransformerMA,
  title={Transformer Memory as a Differentiable Search Index},
  author={Yi Tay and Vinh Q. Tran and Mostafa Dehghani and Jianmo Ni and Dara Bahri and Harsh Mehta and Zhen Qin and Kai Hui and Zhe Zhao and Jai Gupta and Tal Schuster and William W. Cohen and Donald Metzler},
  journal={ArXiv},
  year={2022},
  volume={abs/2202.06991},
  url={https://api.semanticscholar.org/CorpusID:246863488}
}

@article{Aoe1989AnED,
  title={An Efficient Digital Search Algorithm by Using a Double-Array Structure},
  author={Jun-ichi Aoe},
  journal={IEEE Trans. Software Eng.},
  year={1989},
  volume={15},
  pages={1066-1077},
  url={https://api.semanticscholar.org/CorpusID:267894119}
}

@article{Kepner2016MathematicalFO,
  title={Mathematical foundations of the GraphBLAS},
  author={Jeremy Kepner and Peter Aaltonen and David A. Bader and Aydın Buluç and Franz Franchetti and John R. Gilbert and Dylan Hutchison and Manoj Kumar and Andrew Lumsdaine and Henning Meyerhenke and Scott McMillan and Carl Yang and John Douglas Owens and Marcin Zalewski and Timothy G. Mattson and Jos{\'e} E. Moreira},
  journal={2016 IEEE High Performance Extreme Computing Conference (HPEC)},
  year={2016},
  pages={1-9},
  url={https://api.semanticscholar.org/CorpusID:3654505}
}

@article{Dao2022FlashAttentionFA,
  title={FlashAttention: Fast and Memory-Efficient Exact Attention with IO-Awareness},
  author={Tri Dao and Daniel Y. Fu and Stefano Ermon and Atri Rudra and Christopher R\'e},
  journal={ArXiv},
  year={2022},
  volume={abs/2205.14135},
  url={https://api.semanticscholar.org/CorpusID:249151871}
}

@article{He2025PLUMAP,
  title={PLUM: Adapting Pre-trained Language Models for Industrial-scale Generative Recommendations},
  author={Ruining He and Lukasz Heldt and Lichan Hong and Raghunandan Keshavan and Shifan Mao and Nikhil Mehta and Zhengyang Su and Alicia Tsai and Yueqi Wang and Shao-Chuan Wang and Xinyang Yi and Lexi Baugher and Baykal Cakici and Ed H. Chi and Cristos Goodrow and Ningren Han and He Ma and Romer Rosales and Abby Van Soest and Devansh Tandon and Su-Lin Wu and Weilong Yang and Yilin Zheng},
  journal={arXiv preprint arXiv:2510.07784},
  year={2025}
}

@article{Deng2025OneRecUR,
  title={OneRec: Unifying Retrieve and Rank with Generative Recommender and Iterative Preference Alignment},
  author={Jiaxin Deng and Shiyao Wang and Kuo Cai and Lejian Ren and Qigen Hu and Weifeng Ding and Qiang Luo and Guorui Zhou},
  journal={ArXiv},
  year={2025},
  volume={abs/2502.18965},
  url={https://api.semanticscholar.org/CorpusID:276617997}
}

@article{Zhou2025OneRecTR,
  title={OneRec Technical Report},
  author={Guorui Zhou and Jiaxin Deng and Jinghao Zhang and Kuo Cai and Lejian Ren and Qiang Luo and Qianqian Wang and Qigen Hu and Rui Huang and Shiyao Wang and Weifeng Ding and Wuchao Li and Xinchen Luo and Xing-Yao Wang and Zexuan Cheng and Zixing Zhang and Bin Zhang and Bo-Wen Wang and Chao Ma and Cheng-bin Song and Chenhui Wang and Di Wang and Dongxue Meng and Fan Yang and Fang-Peng Zhang and Feng Jiang and Fuxing Zhang and Gang Wang and Guowang Zhang and Han Li and Hengrui Hu and Hezheng Lin and Hongtao Cheng and Hong Cao and Huanjie Wang and Jiaming Huang and Jiapeng Chen and Jiaqian Liu and Jinghui Jia and Kun Gai and Lantao Hu and Liang Zeng and Liao Yu and Qiang Wang and Qidong Zhou and Shengzhe Wang and Shi He and Shuang Yang and Shu-Jun Yang and Sui Huang and Tao Wu and TIAN-CHEN He and Tingting Gao and Wei Yuan and Xiaoyan Liang and Xiao-Xue Xu and Xugang Liu and Yan Wang and Yi Wang and Yiwu Liu and Yue Song and Yufei Zhang and Yun-Hong Wu and Yunfeng Zhao and Zhanyun Liu},
  journal={ArXiv},
  year={2025},
  volume={abs/2506.13695},
  url={https://api.semanticscholar.org/CorpusID:279410085}
}

@article{Zhou2025OneRecV2TR,
  title={OneRec-V2 Technical Report},
  author={Guorui Zhou and Hengrui Hu and Hongtao Cheng and Huanjie Wang and Jiaxin Deng and Jinghao Zhang and Kuo Cai and Lejian Ren and Lu Ren and Liao Yu and Pengfei Zheng and Qiang Luo and Qianqian Wang and Qigen Hu and Rui Huang and Ruiming Tang and Shiyao Wang and Shu-Jun Yang and Tao Wu and Wuchao Li and Xin-Jing LuO and Xing-Yao Wang and Yi Su and Yun-Jie Wu and Zexuan Cheng and Zhanyun Liu and Zixing Zhang and Bin Zhang and Bo-Long Wang and Chao Ma and Cheng-bin Song and Chenhui Wang and Chenglong Chu and Di Wang and Dongxue Meng and Dunju Zang and Fan Yang and Fang-Peng Zhang and Fengqing Jiang and Fuxing Zhang and Gang Wang and Guowang Zhang and Han Li and Honghui Bao and Hong Cao and Jiaming Huang and Jiapeng Chen and Jiaqian Liu and Jinghui Jia and Kun Gai and Lantao Hu and Liang Zeng and Qiang Wang and Qidong Zhou and Rong-Qing Zhang and Shengzhe Wang and Shi He and Shuang Yang and Si-Yuan Mao and Sui Huang and TIAN-CHEN He and Tingting Gao and Wei Yuan and Xiaofeng Liang and Xiao-Xue Xu and Xugang Liu and Yan Wang and Yang Zhou and Yi Wang and Yiwu Liu and Yuexin Song and Yufei Zhang and Yunfeng Zhao and Zhixin Ling and Zi‐Yue Li},
  journal={ArXiv},
  year={2025},
  volume={abs/2508.20900},
  url={https://api.semanticscholar.org/CorpusID:280949827}
}

@article{Liu2025OneRecThinkIR,
  title={OneRec-Think: In-Text Reasoning for Generative Recommendation},
  author={Zhanyun Liu and Shiyao Wang and Xing-Yao Wang and Rong-Qing Zhang and Jiaxin Deng and Honghui Bao and Jinghao Zhang and Wuchao Li and Pengfei Zheng and Xiangyu Wu and Yifei Hu and Qigen Hu and Xinchen Luo and Lejian Ren and Zixing Zhang and Qianqian Wang and Kuo Cai and Yun-Jie Wu and Hongtao Cheng and Zexuan Cheng and Lu Ren and Huanjie Wang and Yi Su and Ruiming Tang and Kun Gai and Guorui Zhou},
  journal={ArXiv},
  year={2025},
  volume={abs/2510.11639},
  url={https://api.semanticscholar.org/CorpusID:282058058}
}

@article{He2016UpsAD,
  title={Ups and Downs: Modeling the Visual Evolution of Fashion Trends with One-Class Collaborative Filtering},
  author={Ruining He and Julian McAuley},
  journal={Proceedings of the 25th International Conference on World Wide Web},
  year={2016},
  url={https://api.semanticscholar.org/CorpusID:1964279}
}

@article{Kamath2025Gemma3T,
  title={Gemma 3 Technical Report},
  author={Gemma Team Aishwarya Kamath and Johan Ferret and Shreya Pathak and Nino Vieillard and Ramona Merhej and Sarah Perrin and Tatiana Matejovicova and Alexandre Ram'e and Morgane Rivi{\`e}re and Louis Rouillard and Thomas Mesnard and Geoffrey Cideron and Jean-Bastien Grill and Sabela Ramos and Edouard Yvinec and Michelle Casbon and Etienne Pot and Ivo Penchev and Gael Liu and Francesco Visin and Kathleen Kenealy and Lucas Beyer and Xiaohai Zhai and Anton Tsitsulin and R{\'o}bert Istvan Busa-Fekete and Alex Feng and Noveen Sachdeva and Benjamin Coleman and Yi Gao and Basil Mustafa and Iain Barr and Emilio Parisotto and David Tian and Matan Eyal and Colin Cherry and Jan-Thorsten Peter and Danila Sinopalnikov and Surya Bhupatiraju and Rishabh Agarwal and Mehran Kazemi and Dan Malkin and Ravin Kumar and David Vilar and Idan Brusilovsky and Jiaming Luo and Andreas Steiner and Abe Friesen and Abhanshu Sharma and Abheesht Sharma and Adi Mayrav Gilady and Adrian Goedeckemeyer and Alaa Saade and Alexander Kolesnikov and Alexei Bendebury and Alvin Abdagic and Amit Vadi and Andr'as Gyorgy and Andr{\'e} Susano Pinto and Anil Das and Ankur Bapna and Antoine Miech and Antoine Yang and Antonia Paterson and Ashish Shenoy and Ayan Chakrabarti and Bilal Piot and Boxi Wu and Bobak Shahriari and Bryce Petrini and Charlie Chen and Charline Le Lan and Christopher A. Choquette-Choo and Cj Carey and Cormac Brick and Daniel Deutsch and Danielle Eisenbud and Dee Cattle and Derek Cheng and Dimitris Paparas and Divyashree Shivakumar Sreepathihalli and Doug Reid and Dustin Tran and Dustin Zelle and Eric Noland and Erwin Huizenga and Eugene Kharitonov and Frederick Liu and Gagik Amirkhanyan and Glenn Cameron and Hadi Hashemi and Hanna Klimczak-Pluci'nska and Harman Singh and Harsh Mehta and Harshal Tushar Lehri and Hussein Hazimeh and Ian Ballantyne and Idan Szpektor and Ivan Nardini and Jean Pouget-Abadie and Jetha Chan and Joe Stanton and J. Michael Wieting and Jonathan Lai and Jordi Orbay and Joe Fernandez and Joshua Newlan and Junsong Ji and Jyotinder Singh and Kat Black and Kathy Yu and Kevin Hui and Kiran Vodrahalli and Klaus Greff and Linhai Qiu and Marcella Valentine and Marina Coelho and Marvin Ritter and Matt Hoffman and Matthew Watson and Mayank Chaturvedi and Michael Moynihan and Min Ma and Nabila Babar and Natasha Noy and Nathan Byrd and Nick Roy and Nikola Momchev and Nilay Chauhan and Oskar Bunyan and Pankil Botarda and Paul Caron and Paul Kishan Rubenstein and Phil Culliton and Philipp Schmid and Pier Giuseppe Sessa and Pingmei Xu and Piotr Stańczyk and Pouya Dehghani Tafti and Rakesh Shivanna and Renjie Wu and Renke Pan and Reza Ardeshir Rokni and Rob Willoughby and Rohith Vallu and Ryan Mullins and Sammy Jerome and Sara Smoot and Sertan Girgin and Shariq Iqbal and Shashir Reddy and Shruti Sheth and Siim P{\~o}der and Sijal Bhatnagar and Sindhu Raghuram Panyam and Sivan Eiger and Susan Zhang and Tianqi Liu and Trevor Yacovone and Tyler Liechty and Uday Kalra and Utku Evci and Vedant Misra and Vincent Roseberry and Vladimir Feinberg and Vlad Kolesnikov and Woohyun Han and Woosuk Kwon and Xi Chen and Yinlam Chow and Yuvein Zhu and Zichuan Wei and Zoltan Egyed and Victor Cotruta and Minh Giang and Phoebe Kirk and Anand Rao and Jessica Lo and Erica Moreira and Luiz Gustavo Martins and Omar Sanseviero and Lucas Gonzalez and Zach Gleicher and Tris Warkentin and Vahab S. Mirrokni and Evan Senter and Eli Collins and Joelle Barral and Zoubin Ghahramani and Raia Hadsell and Yossi Matias and D. Sculley and Slav Petrov and Noah Fiedel and Noam Shazeer and Oriol Vinyals and Jeffrey Dean and Demis Hassabis and Koray Kavukcuoglu and Cl{\'e}ment Farabet and Elena Buchatskaya and Jean-Baptiste Alayrac and Rohan Anil and Dmitry Lepikhin and Sebastian Borgeaud and Olivier Bachem and Armand Joulin and Alek Andreev and Cassidy Hardin and Robert Dadashi and L'eonard Hussenot},
  journal={ArXiv},
  year={2025},
  volume={abs/2503.19786},
  url={https://api.semanticscholar.org/CorpusID:277313563}
}

@article{Yang2024UnifyingGA,
  title={Unifying Generative and Dense Retrieval for Sequential Recommendation},
  author={Liu Yang and Fabian Paischer and Kaveh Hassani and Jiacheng Li and Shuai Shao and Zhang Gabriel Li and Yun He and Xue Feng and Nima Noorshams and Sem Park and Bo Long and Robert Nowak and Xiaoli Gao and Hamid Eghbalzadeh},
  journal={Trans. Mach. Learn. Res.},
  year={2024},
  volume={2025},
  url={https://api.semanticscholar.org/CorpusID:274423313}
}

@article{Jouppi2017IndatacenterPA,
  title={In-datacenter performance analysis of a tensor processing unit},
  author={Norman P. Jouppi and Cliff Young and Nishant Patil and David A. Patterson and Gaurav Agrawal and Raminder Bajwa and Sarah Bates and Suresh Bhatia and Nan Boden and Al Borchers and Rick Boyle and Pierre-luc Cantin and Clifford Chao and Chris Clark and Jeremy Coriell and Mike Daley and Matt Dau and Jeffrey Dean and Ben Gelb and Taraneh Ghaemmaghami and Rajendra Gottipati and William Gulland and Robert Hagmann and C. Richard Ho and Doug Hogberg and John Hu and Robert Hundt and Dan Hurt and Julian Ibarz and Aaron Jaffey and Alek Jaworski and Alexander Kaplan and Harshit Khaitan and Daniel Killebrew and Andy Koch and Naveen Kumar and Steve Lacy and James Laudon and James Law and Diemthu Le and Chris Leary and Zhuyuan Liu and Kyle Lucke and Alan Lundin and Gordon MacKean and Adriana Maggiore and Maire Mahony and Kieran Miller and Rahul Nagarajan and Ravi Narayanaswami and Ray Ni and Kathy Nix and Thomas Norrie and Mark Omernick and Narayana Penukonda and Andy Phelps and Jonathan Ross and Matt Ross and Amir Salek and Emad Samadiani and Chris Severn and Gregory Sizikov and Matthew Snelham and Jed Souter and Dan Steinberg and Andy Swing and Mercedes Tan and Gregory Thorson and Bo Tian and Horia Toma and Erick Tuttle and Vijay Vasudevan and Richard Walter and Walter Wang and Eric Wilcox and Doe Hyun Yoon},
  journal={2017 ACM/IEEE 44th Annual International Symposium on Computer Architecture (ISCA)},
  year={2017},
  pages={1-12},
  url={https://api.semanticscholar.org/CorpusID:4202768}
}

@article{Li2020TheDL,
  title={The Deep Learning Compiler: A Comprehensive Survey},
  author={Mingzhen Li and Yi Liu and Xiaoyan Liu and Qingxiao Sun and Xin You and Hailong Yang and Zhongzhi Luan and Depei Qian},
  journal={IEEE Transactions on Parallel and Distributed Systems},
  year={2020},
  volume={32},
  pages={708-727},
  url={https://api.semanticscholar.org/CorpusID:211069666}
}

@inproceedings{Liao2025EliminatingOR,
  title={Eliminating Out-of-Domain Recommendations in LLM-based Recommender Systems: A Unified View},
  author={Hao Liao and Jiwei Zhang and Jianxun Lian and Wensheng Lu and Mingqi Wu and Shuo Wang and Yong Zhang and Yitian Huang and Mingyang Zhou and Rui Mao},
  year={2025},
  url={https://api.semanticscholar.org/CorpusID:278339501}
}

@article{Linden2003AmazoncomRI,
  title={Amazon.com Recommendations: Item-to-Item Collaborative Filtering},
  author={Greg Linden and Brent Smith and Jeremy York},
  journal={IEEE Internet Comput.},
  year={2003},
  volume={7},
  pages={76-80},
  url={https://api.semanticscholar.org/CorpusID:263872610}
}

@article{Covington2016DeepNN,
  title={Deep Neural Networks for YouTube Recommendations},
  author={Paul Covington and Jay K. Adams and Emre Sargin},
  journal={Proceedings of the 10th ACM Conference on Recommender Systems},
  year={2016},
  url={https://api.semanticscholar.org/CorpusID:207240067}
}

@inproceedings{Vaswani2017AttentionIA,
  title={Attention is All you Need},
  author={Ashish Vaswani and Noam Shazeer and Niki Parmar and Jakob Uszkoreit and Llion Jones and Aidan N. Gomez and Lukasz Kaiser and Illia Polosukhin},
  booktitle={Neural Information Processing Systems},
  year={2017},
  url={https://api.semanticscholar.org/CorpusID:13756489}
}

@software{jax2018github,
  author = {James Bradbury and Roy Frostig and Peter Hawkins and Matthew James Johnson and Chris Leary and Dougal Maclaurin and George Necula and Adam Paszke and Jake Vander{P}las and Skye Wanderman-{M}ilne and Qiao Zhang},
  title = {{JAX}: composable transformations of {P}ython+{N}um{P}y programs},
  url = {http://github.com/jax-ml/jax},
  version = {0.3.13},
  year = {2018},
}

@software{flax2020github,
  author = {Jonathan Heek and Anselm Levskaya and Avital Oliver and Marvin Ritter and Bertrand Rondepierre and Andreas Steiner and Marc van {Z}ee},
  title = {{F}lax: A neural network library and ecosystem for {JAX}},
  url = {http://github.com/google/flax},
  version = {0.12.3},
  year = {2024},
}

@misc{sabne2020xla,
  title={Xla: Compiling machine learning for peak performance},
  author={Sabne, Amit},
  year={2020}
}

@inproceedings{Hong1981IOCT,
  title={I/O complexity: The red-blue pebble game},
  author={Jia-Wei Hong and H. T. Kung},
  booktitle={Symposium on the Theory of Computing},
  year={1981},
  url={https://api.semanticscholar.org/CorpusID:8410593}
}

@article{Hou2022TowardsUS,
  title={Towards Universal Sequence Representation Learning for Recommender Systems},
  author={Yupeng Hou and Shanlei Mu and Wayne Xin Zhao and Yaliang Li and Bolin Ding and Ji-rong Wen},
  journal={Proceedings of the 28th ACM SIGKDD Conference on Knowledge Discovery and Data Mining},
  year={2022},
  url={https://api.semanticscholar.org/CorpusID:249625869}
}

@article{Rajbhandari2019ZeROMO,
  title={ZeRO: Memory optimizations Toward Training Trillion Parameter Models},
  author={Samyam Rajbhandari and Jeff Rasley and Olatunji Ruwase and Yuxiong He},
  journal={SC20: International Conference for High Performance Computing, Networking, Storage and Analysis},
  year={2019},
  pages={1-16},
  url={https://api.semanticscholar.org/CorpusID:269617042}
}


\newpage

\noindent
{\LARGE\textbf{Appendix}}

\appendix

\section{Hardware Optimization Discussions}\label{app:hardware_detailed}
We shall now discuss some of the additional hardware-related details of our constrained decoding implementation.

\subsection{JIT Compilation and Data Structures}
The XLA compilers for TPUs require static tensor shapes at compile time to generate optimized fusion kernels. However, standard trie traversals are inherently dynamic: node $A$ may have $2$ children, while node $B$ may have $1000$. In a CPU environment, one would simply allocate dynamic vectors. But in an XLA context, chasing pointers with dynamic vectors would trigger compilation errors.

Our Vectorized Node Transition Kernel (VNTK) (Algorithm \ref{alg:jax_lookup}) addresses this using static gather operations. Instead of processing a variable number of children, we process precisely $B_\ell$, where this value is the maximum branch factor at level $\ell$. We then use a vectorized \textbf{Gather} operation (e.g., \texttt{jnp.take} with \texttt{mode=`fill'}) to fetch exactly $B_\ell$ potential transitions for every beam in parallel.

To handle nodes with $N_{child} < \mathbf{B}_\ell$ children, we compute a validity mask on the fly ($ m_{valid} = \text{Range}(\mathbf{B}_\ell) < N_{child} $) to zero out invalid entries in the score calculation. This transformation converts a dynamic control flow problem (``loop $N$ times'') into a static data flow problem (``gather fixed $\mathbf{B}_\ell$ elements and mask''). This ensures the entire decoding step remains a single, static XLA graph, enabling full loop unrolling and pipelining on TPU systolic arrays.

\subsubsection{Stacked CSR Layout}
To further minimize memory access latency of VNTK, we employ a stacked CSR memory layout. In a standard CSR format, column indices and data values are stored in separate arrays, typically requiring two memory fetches per transition. In STATIC, we stack these into a single $(N_{edges}, 2)$ tensor. This allows coalesced reads to retrieve both the candidate token ID and the pointer to the next node in a single memory transaction, effectively halving the number of random memory accesses.

\subsubsection{Dense Mask Optimization}
While sparse traversal is efficient for deep layers of the trie, the initial layers exhibit very high branching factors. 
To address this, for the first $d$ decoding steps (where $0 \le d < L$), we bypass the VNTK and utilize a pre-computed dense boolean mask tensor to represent all valid prefixes of length $d$. This hybrid approach allows us to trade a small amount of static memory for a significant throughput gain in the most computationally expensive layers of the retrieval process. In our experiments in Section \ref{sec:experiment_yt}, we set $d=2$ to densify the first two layers of the trie.

\subsubsection{Variable Length SIDs}

Additionally, we want to highlight that while our evaluation focused on fixed-length SIDs with a fixed-size discrete codebook, the CSR-based transition matrix and the VNTK kernel are inherently compatible with variable-length sequences with variable-size codebooks. This is because a transition can be drawn to a null node at any level of the prefix tree and because the branch factor does not have to be fixed so long as we have an index assigned to every node, which is a requirement of the CSR format.

\subsection{Cross-Platform Portability (JAX \& PyTorch)}
While our primary implementation utilizes JAX for XLA compilation, STATIC is inherently portable to PyTorch and CUDA environments. The shift to a vectorized \textbf{Gather}-based kernel unifies the implementation strategy across frameworks. In JAX/XLA, the kernel relies on \texttt{jnp.take} (Gather). In PyTorch/Inductor, this logic translates directly to \texttt{torch.gather}.

    

\section{Analysis of STATIC Memory Usage}
\label{app:static_memory_analysis}

We present an analysis of the HBM memory usage of STATIC. 

\subsection{Absolute Upper Bound}
The core memory usage for STATIC lies in the storage of the associated trie (which represents the restricted vocabulary) and the dense tensor mask $\mathbf{D}$. For the dense tensor mask, the storage cost is simply the cost of storing $|\mathcal{V}|^d$ bits and $|\mathcal{V}|^d$ state IDs. The total storage cost for a trie in CSR format is a function of the number of unique prefix nodes at each level $\ell$ of the tree. For a restricted vocabulary of $|\mathcal{C}|$ items, where each item is a Semantic ID (SID) of length $L$ and each token has a vocabulary size $|\mathcal{V}|$, the number of unique nodes at level $\ell$ ($N_\ell$) is strictly limited by two factors:
\begin{enumerate}
    \item \textbf{Level Capacity}: The maximum possible unique sequences of length $\ell$, which is $|\mathcal{V}|^\ell$.
    \item \textbf{Total Constraints}: The total number of unique constraint items $|\mathcal{C}|$, as no level can have more unique prefixes than there are total items in the vocabulary.
\end{enumerate}

The absolute upper bound for nodes at level $\ell$ is therefore:
\[
N_\ell \leq \min(|\mathcal{V}|^\ell, |\mathcal{C}|)
\]

The total storage upper bound in bytes, $U_{\text{max}}(K_1, K_2, |\mathcal{V}|, |\mathcal{C}|, L, d)$, is given by a contribution of $(\frac{1}{8}+K_2) |\mathcal{V}|^d$ for the dense tensor mask ($1$ bit for the mask and $K_2$ bytes per state, where $K_2=4$ typically), as well as by multiplying the sum of the per-level prefix node upper bounds for later levels by the storage cost per node $K_1$ (typically 12 bytes, due to the three arrays of the CSR sparse matrix format):
\[
U_{\text{max}}(K_1, K_2, |\mathcal{V}|, |\mathcal{C}|, L, d) = \left(\frac{1}{8} +K_2\right) |\mathcal{V}|^d + K_1 \cdot \sum_{\ell=d+1}^{L} \min(|\mathcal{V}|^\ell, |\mathcal{C}|)
\]

\subsection{Calculation for YouTube}
\label{sec:calc_for_yt}
We compute the upper bound for the YouTube setting used in Section \ref{sec:experiment_yt}. Our main configuration variables are fixed to $|\mathcal{V}|=2048$, $L=8$, $K_1=12$, $K_2=4$, and $d=2$. Note since we have $20$ million constraints in the restricted vocabulary, we also have $|\mathcal{C}|=20 \cdot 10^6$. Now we compute the total memory usage
\begin{enumerate}
    \item \textbf{Dense Mask Phase ($\ell=1, 2$):} 
    \begin{itemize}
        \item The number of bytes we need to store the dense mask is simply $\left(\frac{1}{8} + 4 \right) \cdot 2048^2 = 17301504$ bytes, or $\approx 17.3$ MB.
    \end{itemize}
    \item \textbf{Constraint Phase ($\ell \ge 3$):} 
    \begin{itemize}
        \item For $3 \leq \ell \leq 8$, the capacity $|V|^n$ is significantly larger than $|\mathcal{C}|$ (e.g., $|V|^3 \approx 8.5$ billion). Hence, each of these $6$ levels contributes $|\mathcal{C}| \cdot 12$ bytes.
        \item Contribution: $(L-2) \cdot |\mathcal{C}| \cdot 12 = 6 \cdot 20 \cdot 10^6 \cdot 12 = 1.44$ GB.
    \end{itemize}
\end{enumerate}

Our maximum per-chip usage is thus $\approx\!1.46\:\text{GB}$, validating our claim of approximately $1.5$ GB of HBM usage for $20$ million constraints. Due to the fact that the YouTube video distribution is highly non-uniform and exhibits significant clustering \cite{He2025PLUMAP}, the semantic ID distribution will also, which produces a significant amount of collision in the prefix space. As such, for most restrict vocabularies, the actual HBM usage will be much less than this upper bound. In practice, we note that our utilization is $\leq\!75\%$ of this amount.

\subsection{Capacity Planning}
In our YouTube setting (Section \ref{sec:experiment_yt}), we use the rule of thumb that every 1 million constraints introduce an additional $\approx 90$ MB of HBM usage. This is because the storage cost is dominated by the constraint phase described in the previous section.

As shown above, the dense mask phase (early levels) contribution remains constant (for fixed $d$), while each additional level in the constraint phase adds exactly $|\mathcal{C}| \times K_1$ bytes to the storage cost.

When $|\mathcal{C}|$ is 1 million:
\begin{itemize}
    \item \textbf{Dense Mask Phase Contribution}: Only $\left(\frac{1}{8} + 4 \right) \cdot 2048^2 = 17301504$ bytes, or $\approx 17.3$ MB.
    \item \textbf{Constraint Phase Contribution}: For $|\mathcal{C}|=10^6$, we see the total contribution is $6 \cdot 10^6 \cdot 12\:\text{bytes}= 72$MB.
    \item \textbf{Total}: $17.3\:\text{MB} + 72\:\text{MB} \approx 90\:\text{MB}$ per million items.
\end{itemize}

For the production-scale restrict corpus in Appendix \ref{sec:calc_for_yt} of $20$ million items, we saw the transition to the constraint phase happen later, resulting in an average of $\approx\!\mathbf{73 \:\text{MB}}$ per 1 million items, making $90$ MB a highly reliable linear upper bound approximation for capacity planning. A similar rule-of-thumb can be computed for different values of $L$, $|\mathcal{V}|$, and $|\mathcal{C}|$ using the principles above.

\subsection{Memory Footprint in Implementation}
While LLM parameters are typically sharded across devices in large-scale serving setups \cite{Rajbhandari2019ZeROMO}, the relatively compact size of the sparse transition matrix in STATIC (e.g. $1.5$ GB memory for $20$ million items) allows a replication strategy where the sparse transition matrix is replicated on the HBM (High-Bandwidth Memory) of every device. This eliminates the need for cross-chip communication (All-Gather or All-Reduce) during the constraint check. Each chip independently computes the validity mask for its local batch of beams, preserving the linear scalability of the serving system.

As item corpora scale toward billions, future research should explore hierarchical storing or sharding strategies to maintain this linear scalability without saturating device memory.

\section{Finite State Transducer (FST) Comparison}
\label{app:fst}

In this section we discuss why we excluded FSTs from our list of baselines in Section \ref{sec:yt_system_efficiency_analysis}. While FSTs are a standard tool for grammar-constrained decoding in NLP, they are fundamentally ill-suited for the scale and hardware requirements of modern generative retrieval. In generative retrieval, each item is represented by a unique Semantic ID (SID) of length L. For a corpus of $|\mathcal{C}|$ items, the number of states (prefix nodes) $N_\ell$ at each level $\ell$ of the search tree is given by $N_\ell \leq \min(|\mathcal{V}|^\ell, |C|)$. For our YouTube deployment with $|\mathcal{C}|=20 \cdot 10^6$ items, a vocabulary size $|\mathcal{V}|=2048$, and $L=8$, the total state count is approximately 124 million nodes.

The feasibility of on-device decoding depends on the memory footprint per state. Under a standard OpenFst VectorFst implementation using the StdArc structure, each arc requires 16 bytes to store input/output labels, a weight, and a next-state ID. Additionally, each state incurs a container overhead (typically 24 bytes for a std::vector in C++). For a trie representing 20 million items with 124 million states, this implies a total memory footprint of 4.96 GB ($124 \cdot 10^6 \cdot 40$ bytes). In contrast, STATIC employs a stacked CSR layout that stores Token ID and Next Node pairs as a contiguous tensor. This requires exactly 12 bytes per node, resulting in a total footprint of 1.5 GB for 20 million items. The pointer-based FST implementation requires over 3x the memory of STATIC.

Even if an FST fits into memory, it is fundamentally hostile to accelerators for two reasons. First, FSTs rely on recursive pointer chasing, which prevents memory coalescing and nullifies hardware prefetchers. Second, standard FST traversals involve data-dependent control flow that triggers costly graph recompilations in ML compilers like XLA. Our CPU Trie baseline shows that this pointer-chasing approach adds 31.3 ms of latency per step, while STATIC adds only 0.03 ms.

\section{Detailed Latency Analysis}
\label{app:latency_scaling}

In this section, we present a fine-grained latency overhead analysis of our STATIC method in the setting of Section \ref{sec:experiment_yt} compared to the baselines in Table \ref{tab:latency_bench_yt}. All numbers given represent the per-step latency throughout the full decoding (over $L=8$ decoding steps). All values represent the additional latency (in milliseconds) incurred by the constraint enforcement logic relative to the unconstrained baseline (which performs no masking), calculated over $100$ trials.

Table \ref{tab:per_step_uptick_stats}a provides the precise numbers (means and standard deviations) for Figure \ref{fig:scaling_constraints}, and the $N=2\cdot 10^7$ row gives the means and standard deviations for Table \ref{tab:latency_bench_yt}. The precise numbers (means and standard deviations) for Figure \ref{fig:scaling_vocab_size} are given in Table \ref{tab:per_step_uptick_stats}b.

\begin{table}[h]
\centering
\caption{Per-Step Latency Overhead (ms) over the Unconstrained Baseline. We report mean and standard deviation over $100$ trials.}
\label{tab:per_step_uptick_stats}
\resizebox{\columnwidth}{!}{%
\begin{tabular}{lccccc}
\toprule
\textbf{Setting} & \textbf{STATIC (Ours)} & \textbf{Hash Bitmap} & \textbf{PPV Approximate} & \textbf{PPV Exact} & \textbf{CPU Trie} \\
\midrule
\multicolumn{6}{c}{\textit{(a) Scaling with Constraint Size $|\mathcal{C}|$ ($|\mathcal{V}|=2048$)}} \\
\midrule
$|\mathcal{C}|=10^5$ & $0.023 \pm 0.008$ & $12.335 \pm 0.020$ & $0.588 \pm 0.008$ & $6.375 \pm 0.268$ & $27.070 \pm 0.416$ \\
$|\mathcal{C}|=10^6$ & $0.030 \pm 0.008$ & $12.339 \pm 0.016$ & $0.846 \pm 0.008$ & $15.368 \pm 0.338$ & $27.579 \pm 0.353$ \\
$|\mathcal{C}|=10^7$ & $0.034 \pm 0.006$ & $12.341 \pm 0.026$ & $1.486 \pm 0.020$ & $31.765 \pm 2.150$ & $28.060 \pm 0.396$ \\
$|\mathcal{C}|=2 \cdot 10^7$ & $0.033 \pm 0.008$ & $12.343 \pm 0.020$ & $1.564 \pm 0.033$ & $34.116 \pm 2.118$ & $31.336 \pm 0.479$ \\
$|\mathcal{C}|=10^8$ & $0.039 \pm 0.010$ & $12.331 \pm 0.013$ & $1.713 \pm 0.038$ & $38.691 \pm 2.469$ & \textit{OOM} \\
\midrule
\multicolumn{6}{c}{\textit{(b) Scaling with Vocabulary Size $|\mathcal{V}|$ ($|\mathcal{C}|=10^7$)}} \\
\midrule
$|\mathcal{V}|=256$ & $0.040 \pm 0.008$ & $1.554 \pm 0.008$ & $1.584 \pm 0.035$ & $6.004 \pm 0.228$ & $27.359 \pm 0.365$ \\
$|\mathcal{V}|=512$ & $0.041 \pm 0.008$ & $3.094 \pm 0.008$ & $1.543 \pm 0.035$ & $10.586 \pm 0.655$ & $27.853 \pm 0.294$ \\
$|\mathcal{V}|=1024$ & $0.038 \pm 0.008$ & $6.180 \pm 0.011$ & $1.509 \pm 0.036$ & $18.360 \pm 1.313$ & $29.239 \pm 0.491$ \\
$|\mathcal{V}|=2048$ & $0.034 \pm 0.006$ & $12.341 \pm 0.026$ & $1.486 \pm 0.020$ & $31.765 \pm 2.150$ & $31.176 \pm 0.446$ \\
$|\mathcal{V}|=4096$ & $0.034 \pm 0.008$ & $24.620 \pm 0.033$ & $1.620 \pm 0.025$ & $57.066 \pm 3.540$ & $29.810 \pm 0.426$ \\
$|\mathcal{V}|=8192$ & $0.041 \pm 0.008$ & $49.191 \pm 0.090$ & $1.668 \pm 0.024$ & $107.015 \pm 9.269$ & $29.383 \pm 0.741$ \\
$|\mathcal{V}|=16384$ & $0.044 \pm 0.008$ & $98.273 \pm 0.025$ & $1.824 \pm 0.031$ & $287.181 \pm 4.321$ & $30.885 \pm 0.276$ \\
$|\mathcal{V}|=32768$ & $0.041 \pm 0.006$ & $196.474 \pm 0.038$ & $2.093 \pm 0.029$ & $578.738 \pm 14.596$ & $31.676 \pm 0.399$ \\
\bottomrule
\end{tabular}%
}
\end{table}

\section{Hardware Scaling with High Branching Factor}
\label{app:appendix_branch_scaling}

To isolate the hardware-level scaling efficiency of the STATIC masking kernel, we design a benchmark that stresses the vectorized burst-read mechanism under extreme sparsity and high-cardinality conditions. Specifically, we evaluate the execution time of the core Vectorized Node Transition Kernel (VNTK), i.e. Algorithm \ref{alg:jax_lookup}, as the maximum branching factor increases.

For each tested max branching factor $B \in \{2^1, 2^2, \dots, 2^{18}\}$, we dynamically set the token vocabulary size such that $|\mathcal{V}| = B$. Under this configuration, we synthesize a new constraint set comprising $|\mathcal{C}|=10^6$ Semantic IDs generated uniformly at random. We then flatten this newly generated prefix tree into a Compressed Sparse Row (CSR) matrix for the VNTK algorithm.

\begin{figure}[t]
  \centering
  \includegraphics[width=\linewidth]{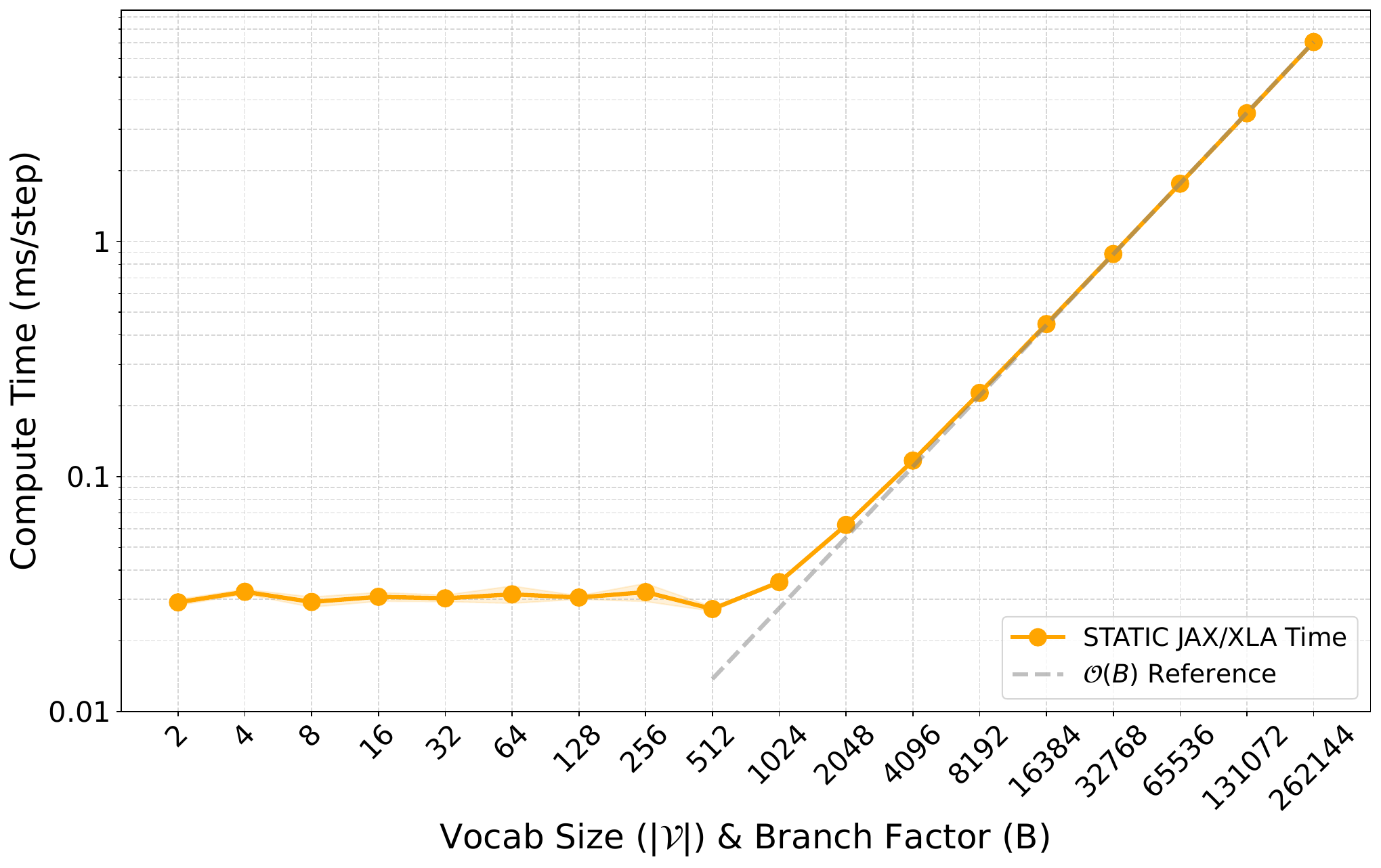}
  \caption{Scaling of the STATIC masking kernel with respect to the max branch factor (log-log scale). We plot means and standard deviations (shaded region) over $100$ trials. For each trial, the token vocabulary size is set to the branch factor, while the number of constraints is fixed at $|\mathcal{C}|=10^6$. The STATIC method exhibits asymptotically linear $\mathcal{O}(B)$ scaling.}
  \label{fig:scaling_mask_kernel}
\end{figure}

During the timed trials, we execute the JIT-compiled masking kernel, which performs a coalesced read of $B$ contiguous elements from HBM and applies the validity mask. To ensure robust measurements, we perform synchronous wall-clock timing over multiple iterations, capturing the amortized pure hardware compute time per operation while aggressively clearing the accelerator memory between scales to prevent out-of-memory (OOM) errors. The scaling results are illustrated in Figure~\ref{fig:scaling_mask_kernel}. Crucially, the benchmark demonstrates that STATIC exhibits strict asymptotic linear scaling, $O(B)$, with respect to the maximum branch factor $B$. We note that STATIC exhibits constant runtime before reaching TPU VMEM bandwidth, at which point the linear regime begins. 

\section{Cold-Start Experiment Details}
\label{appendix:cold_start}

Here, we add additional details of the evaluation protocol and dense retrieval baseline for the cold-start experiments from Section \ref{sec:cold_start_amazon}.

The evaluation protocol for the Amazon dataset is as follows. For each item history sequence in the cold-start split, the model generates output items either through autoregressive decoding of Semantic IDs or by dense retrieval. We report the peak Recall@1 results achieved over $70$ epochs of training, which we found sufficient for convergence across the tested subdatasets.


The dense retrieval baseline is a two-tower model designed to align sequential user behavior with static item content features. The ``query tower'' utilizes the Gemma 1B architecture to convert the sequence of Semantic IDs into a 128-dimensional query vector using a projection head. The ``item tower'' is a Multi-Layer Perceptron (MLP) with hidden layer dimensions of ($512$, $256$) and interspersed dropout layers that have a dropout rate of $0.1$. This tower ingests precomputed 768-dimensional T5 item embeddings \cite{Raffel2019ExploringTL} to obtain an 128-dimensional item representation. The dense model is optimized using an in-batch contrastive loss with a batch size of 512 to maximize the similarity between query and ground-truth item vectors. The constrained recommendation output is generated by Maximum Inner Product Search (MIPS) with only the cold-start item vectors, for a fair comparison with other cold-start techniques.

\section{STATIC Code: JAX Implementation}

In this section, we give a code snippet to demonstrate an explicit example implementation of Algorithms \ref{alg:constrained_decoding} and \ref{alg:jax_lookup} using the JAX \cite{jax2018github} and Flax \cite{flax2020github} frameworks. Note that the code works on both TPU and GPU with almost identical latency overhead.

\begin{minted}[
    frame=lines,
    framesep=2mm,
    baselinestretch=1.2,
    fontsize=\footnotesize,
]{python}
# Copyright 2026 Google LLC.
# SPDX-License-Identifier: Apache-2.0
"""Implementation of Hardware-Accelerated Constrained Decoding algorithms."""

import functools
import flax
import jax
import jax.numpy as jnp

NEG_INF = -1.0e10


@flax.struct.dataclass(frozen=True)
class TransitionMatrix:
  """CSR-based Transition Matrix with optional dense optimizations."""

  row_pointers: jnp.ndarray
  data: jnp.ndarray
  start_mask_packed: jnp.ndarray | None = None
  l1_dense_mask_packed: jnp.ndarray | None = None
  l1_dense_states: jnp.ndarray | None = None


def vectorized_sparse_candidate_extraction(
    log_probs, nodes, transition_matrix, layer_max_branches, vocab_size
):
  """Algorithm 2: Vectorized Node Transition Kernel (VNTK)."""

  n_flat, lp_flat = nodes.reshape(-1), log_probs.reshape(-1, vocab_size)
  starts = transition_matrix.row_pointers[n_flat]
  lens = transition_matrix.row_pointers[n_flat + 1] - starts
  offsets = jnp.arange(layer_max_branches)
  gathered = jnp.take(
      transition_matrix.data,
      starts[:, None] + offsets[None, :],
      axis=0,
      mode="fill",
      fill_value=0,
  )
  valid_mask = offsets[None, :] < lens[:, None]
  dense_indices, dense_data = gathered[:, :, 0], gathered[:, :, 1]
  dense_data = jnp.where(
      valid_mask, dense_data, 0
  )
  candidate_lp = jnp.take_along_axis(
      lp_flat, jnp.clip(dense_indices, max=vocab_size - 1), axis=1
  )
  return (
      jnp.where(valid_mask, candidate_lp, NEG_INF),
      dense_indices,
      dense_data,
      valid_mask,
  )


@functools.partial(
    jax.jit, static_argnames=("vocab_size", "layer_max_branches", "step")
)
def hardware_accelerated_constrained_decoding_step(
    logits,
    previous_nodes,
    transition_matrix,
    layer_max_branches,
    vocab_size,
    step=-1,
):
  """Algorithm 1: Hardware-Accelerated Constrained Decoding Step."""
  log_probs = jax.nn.log_softmax(logits)

  if step == 0 and transition_matrix.start_mask_packed is not None:
    mask = jnp.unpackbits(transition_matrix.start_mask_packed).astype(bool)[
        :vocab_size
    ]
    return jnp.where(mask, log_probs, NEG_INF), jnp.broadcast_to(
        jnp.arange(vocab_size, dtype=jnp.int32) + 1, logits.shape
    )
  if step == 1 and transition_matrix.l1_dense_mask_packed is not None:
    parents = previous_nodes - 1
    mask = jnp.unpackbits(
        transition_matrix.l1_dense_mask_packed[parents], axis=-1
    )
    return (
        jnp.where(mask.astype(bool)[..., :vocab_size], log_probs, NEG_INF),
        transition_matrix.l1_dense_states[parents],
    )
  sparse_lp, sparse_toks, sparse_next, valid_mask = (
      vectorized_sparse_candidate_extraction(
          log_probs,
          previous_nodes,
          transition_matrix,
          layer_max_branches,
          vocab_size,
      )
  )
  scatter_idx = jnp.where(valid_mask, sparse_toks, vocab_size)
  masked_lp = jnp.full((previous_nodes.size, vocab_size + 1), NEG_INF)
  masked_lp = masked_lp.at[
      jnp.arange(previous_nodes.size)[:, None], scatter_idx
  ].set(sparse_lp)

  return (
      masked_lp[:, :vocab_size].reshape(logits.shape),
      sparse_next.reshape(previous_nodes.shape + (layer_max_branches,)),
  )
\end{minted}

\end{document}